\documentclass[aps,groupedaddress]{revtex4}
\usepackage{epsfig}
\usepackage{graphicx}
\usepackage[T1]{fontenc}
\usepackage{ae}
\usepackage{color}
\usepackage[latin1]{inputenc}
\usepackage{amssymb,amsbsy,amsmath}
\usepackage{bbm}
\newcommand{\op}[1]{\fontdimen12\textfont3=2pt\fontdimen12\scriptfont3=1.4pt\!\null\mathop{\protect\vphantom{#1}\smash{#1}}\limits_{\sim}\null\!}

\newtheorem{lemma}{Lemma}

\newtheorem{prop}{Proposition}

\newtheorem{theorem}{Theorem}

\begin{document}


\title[Gibbs states]
      {Classical limit of Gibbs states for quantum spin systems}
\author{Heinz-J\"urgen Schmidt
}
\address{ Universit\"at Osnabr\"uck,
Fachbereich Physik,
 D - 49069 Osnabr\"uck, Germany}


\begin{abstract}
We study the relation between quantum mechanical and classical Gibbs states of spin systems with spin quantum number $s$.
It is known that quantum states and observables can be represented by functions defined on the phase space ${\mathcal S}$, which in our case is the $N$-fold product of unit spheres. Therefore, the classical limit $s\to\infty$ of (suitably scaled) quantum Gibbs states can be described as the limit of functions defined on ${\mathcal S}$. We choose to approximate the exponential function of the Hamiltonian by a polynomial of degree $n$ and thus have to deal with the problem of the limit of double sequences (depending on $n$ and $s$) treated in the theorem of Moore-Osgood.
The convergence of quantum Gibbs states to classical ones is illustrated by the example of the Heisenberg dimer.
We apply our method to the explicit calculation of the phase space function describing spin monomials,
and finally add some general remarks on the theory of spin coherent states.
\end{abstract}

\maketitle

\section{Introduction}\label{sec:I}

The classical limit of quantum theory is an essentially unsolved problem, despite partial results,
only a small selection of which we cite here \cite{D45,VC89,W95,V20,C21}.
A fundamental problem is to formulate two conceptually distinct theories within a common framework.
The core of such a framework could be the statistical duality based on the notions of ``states", ``effects" and ``probability", see \cite{L12}.
Another difficulty is that the formal limit $\hbar \rightarrow 0$ is, strictly speaking, meaningless, since the value of $\hbar$
depends on the chosen system of physical units, but see \cite{WW95}.
For the subproblem of the classical limit of spin systems, this difficulty vanishes:
We can choose the units such that $\hbar=1$, and are instead concerned with the limit $s\rightarrow\infty$, where $s={\textstyle \frac{1}{2}},1,{\textstyle \frac{3}{2}},\ldots$ denotes the dimensionless spin quantum number.
The classical limit of spin systems is relevant not only conceptually but also for practical reasons: Physical models that hold for different values of $s$ are often checked for consistency in the classical limit of $s\rightarrow\infty$.
Sometimes the classical treatment is a surprisingly good approximation. For magnetic molecules, the spin quantum number of gadolinium $s={\textstyle \frac{7}{2}}$ is already close to $s=\infty$ for practical purposes due to the seven unpaired electrons in the $f$ shell, see \cite{P15}.

The aforementioned problem of how to close the conceptual gap between classical and quantum theory has also been solved for spin systems.
Using the tool of ``spin coherent states'' \cite{P72,L73,B75},
general states and observables living in different Hilbert spaces for different $s$ can be represented by functions on one and the same classical phase space ${\mathcal S}$. More precisely, observables $A$ can be represented by {\em contravariant symbols} $G(A)$ and (pure or mixed) states $W$ can be represented by {\em covariant symbols} $g(W)$, so that the well-known statistical duality
expressed by the formula $\mbox{Tr } W A$ for the expected value is exactly reproduced by the integral of the product of $G(A)$ and $g(W)$. Recall that the mappings $g$ or $G$ generally do not map products of matrices to products of functions, and therefore the quantum character of spin theory is preserved despite this classical-looking representation.

This representation thus allows us to speak of limits of quantum objects for $s\rightarrow\infty$, since these limits can be understood as limits of functions defined on the classical phase space. The first results on the classical limit of partition functions and ground-state energies can be found in the seminal paper \cite{L73}, although the formulation is somewhat different from our work. For example, the Berezin-Lieb inequalities (4.12) and (5.6) in \cite{L73} are not
formulated as limit statements. The next work that extends these results to the classical limit of time evolution is \cite{FKL07}. This brilliant work has a small flaw due to the wrong factors in the definition of spin-coherent states and in the following equation (19), but as far as I can see this has no consequences for the main results.

Of course, there are still some open issues: one is the classical limit of Gibbs states $\exp\left( -\beta \widehat{H}^{(s)}\right)/Z$ characterizing equilibrium states due to a heat bath with an inverse temperature $\beta$.
This problem is addressed in the present work. Based on the aforementioned preliminary work, it is almost clear in what sense this limit should be understood. First, Gibbs states are states and convergence should be studied for their covariant symbols. Second, the energy of, say, a system of $N$ spins with Heisenberg coupling scales with $s^2$ and should be appropriately rescaled before considering the classical limit. Moreover, a second rescaling of the partition function by inserting the factor $(2s+1)^{-N}$, see \cite{L73}, eq.~(3.1), is crucial for convergence to a classical limit and should therefore also be considered for the Gibbs states.
The next question is about the appropriate topology for which convergence can be proved. We are always dealing with {\em continuous} functions
on the compact phase space ${\mathcal S}$ and therefore the so-called sup-norm lends itself since the space ${\mathcal C}\left( {\mathcal S}\right)$ of continuous functions on ${\mathcal S}$ will be complete with respect to the sup-norm.

After recalling the general definitions in Section \ref{sec:D}, we address the problem of the classical limit of Gibbs states, see Section \ref{sec:G}.
The covariant symbols of the products of scaled spin operators have a natural classical limit consisting of the products of the corresponding
components of the classical spin vectors, see \cite{L73} and \cite{FKL07}. The corresponding  result is reformulated and proved in a context suitable for the present purposes, see lemma \ref{LemmaProduct} in section \ref{sec:GE}. It implies the convergence of the covariant symbols of the polynomials of the
Hamiltonian $\widehat{{H}}^{(s)}$ to their classical counterpart for $s \to\infty$. On the other hand, the $\exp(-\beta \widehat{{H}}^{(s)})$-part of the Gibbs state is,
in the obvious sense, the limit of polynomials (truncated Taylor series) $W(N,s,n)$ of degree $n$ for $n\to \infty$. Thus, we face the problem that the iterated limits of a double series as a function of $s$ and $n$ need not coincide. A sufficient condition for this coincidence is the uniform convergence of one of these double series (Moore-Osgood theorem).
We apply this theorem by showing that the classical limit $s \to\infty$ of the covariant symbols of the polynomials $W(N,s,n)$ exists and is uniform in $n$. Together with the classical limit of the partition function, we thus arrive at our main result in Theorem \ref{TGG}, Section \ref{sec:GG}.
The simplest example is the Gibbs state of the Heisenberg dimer, which is used for illustration in section \ref{sec:HD}.

In section \ref{sec:A} we will apply the ideas of the proof of the main theorem to obtain closed expressions for the covariant symbol of spin monomials.
Finally, section \ref{sec:CS} will be devoted to some general, 
mainly pedagogical remarks on the geometry of coherent spin states and on statistical duality.
We close with a Summary in Section \ref{sec:SUM}.

\section{Definitions and first results}\label{sec:D}

Let $s$ be the integer or half-integer spin quantum number and ${\mathcal H}^{(1,s)}$ the corresponding $2s+1$-dimensional Hilbert space
that carries an irreducible unitary representation $U_g,\,g\in SU(2),$ with Hermitean generators $\op{s}_i,\,i=1,2,3$, also called
``spin operators". W.~r.~t.~the eigenbasis $|m\rangle,\,m=s,\ldots,-s$ of  $\op{s}_3$ the Hilbert space ${\mathcal H}^{(1,s)}$ can be
identified with ${\mathbbm C}^{2s+1}$.
As usual, one defines the raising/lowering operators by $\op{s}_\pm := \op{s}_1\pm  {\sf i} \op{s}_2$.
The scaled operators will be denoted by a hat, e.~g., $\widehat{\op{s}}_i := \frac{1}{s}\,\op{s}_i$ for $i=1,2,3$.
The explicit form of the spin operators w.~r.~t.~the $|m\rangle$-basis can be found in any textbook on quantum mechanics,
e.~g.~, we have
\begin{equation}\label{Dspin}
 \op{s}_\pm \,\left| m \right\rangle = \sqrt{s(s+1)-m(m\pm 1)}\,\left| m \pm 1\right\rangle
 \;,
\end{equation}
for $m=-s,\ldots,s$, where the square root in (\ref{Dspin}) vanishes if $\left| m \pm 1\right\rangle$ is not defined.

Let ${\mathbbm S}^2$ denote the unit sphere in ${\mathbb R}^3$ and
$\Omega\in {\mathbbm S}^2$  an arbitrary unit vector parametrized by spherical coordinates $(\theta, \varphi)$ such that
\begin{equation}\label{D1}
 \Omega=\left(
 \begin{array}{l}
   \sin \theta\, \cos\varphi \\
   \sin \theta\, \sin\varphi \\
   \cos\theta
 \end{array}
 \right)\;,
\end{equation}
where $0\le \theta \le \pi$ and $0\le \varphi <2\,\pi$ (not defined for $\theta=0,\pi$).
${\mathbbm S}^2$ may be viewed as the classical single spin phase space equipped with its natural symplectic form.
We will define the special unitaries
\begin{equation}\label{D2}
 U(\Omega):= \exp\left[ {\scriptsize\frac{\theta}{2}} \left( e^{{\sf i}\varphi}\op{s}_-  -  e^{-{\sf i}\varphi}\op{s}_+\right)\right]
 \;,
\end{equation}
for all $\Omega\in  {\mathbbm S}^2$ (except for  $\theta=0,\pi$)
and the correspondingly transformed spin operators
\begin{equation}\label{D3}
  \op{s}_i(\Omega):= U^\ast(\Omega)\,\op{s}_i\,U(\Omega)
  =: s_{i3}(\Omega)\,\op{s}_3 +s_{i-}(\Omega)\,\op{s}_- + s_{i+}(\Omega)\,\op{s}_+
  ,\;i=1,2,3
  \;.
\end{equation}
In this respect we follow \cite{FKL07}, but we have to correct the equation (19) of this reference for the explicit form of $\op{s}_i(\Omega)$.
After some calculations we obtain
\begin{eqnarray}
\label{D4a}
   \op{s}_1(\Omega) &=& \sin\theta\,\cos\varphi\,\op{s}_3
    +{\scriptsize\frac{1}{2}}\,e^{{\sf i}\varphi}\left( \cos\theta\cos\varphi-{\sf i}\sin\varphi\right) \op{s}_-
   +{\scriptsize\frac{1}{2}}\,e^{-{\sf i}\varphi}\left( \cos\theta\cos\varphi+{\sf i}\sin\varphi\right) \op{s}_+\;, \\
   \label{D4b}
   \op{s}_2(\Omega) &=& \sin\theta\,\sin\varphi\,\op{s}_3
    +{\scriptsize\frac{1}{2}}\,e^{{\sf i}\varphi}\left( \cos\theta\sin\varphi+{\sf i}\cos\varphi\right) \op{s}_-
   +{\scriptsize\frac{1}{2}}\,e^{-{\sf i}\varphi}\left( \cos\theta\sin\varphi-{\sf i}\cos\varphi\right) \op{s}_+ \;,\\
   \label{D4c}
  \op{s}_3(\Omega) &=& \cos\theta\,\op{s}_3 -{\scriptsize\frac{1}{2}}\,e^{{\sf i}\varphi}\,\sin\theta\,\op{s}_-
  -{\scriptsize\frac{1}{2}}\,e^{-{\sf i}\varphi}\,\sin\theta\,\op{s}_+
  \;.
\end{eqnarray}

For later use we note the following
\begin{lemma}\label{LemmaComp}
 For all $i=1,2,3$ there holds
 \begin{equation}\label{DC1}
   \left| s_{i,\pm}(\Omega)\right| \le {\scriptsize\frac{1}{2}}
   \;,
 \end{equation}
 and
  \begin{equation}\label{DC2}
   \left| s_{i,3}(\Omega)\right| \le 1
   \;.
 \end{equation}
\end{lemma}
The {\bf proof} is straightforward. For example,
\begin{eqnarray}
\label{DC3a}
   \left| s_{1,-}(\Omega)\right|  &\stackrel{(\ref{D4a})}{=}& {\scriptsize\frac{1}{2}}\left|
   e^{{\sf i}\varphi}\left( \cos\theta\cos\varphi-{\sf i}\sin\varphi\right)
   \right|
    \\
   &=& {\scriptsize\frac{1}{2}}\sqrt{\cos^2\theta\, \cos^2\varphi+\sin^2\varphi}
    \\
  &\le& {\scriptsize\frac{1}{2}}\sqrt{\cos^2\varphi+\sin^2\varphi}={\scriptsize\frac{1}{2}}
  \;.
\end{eqnarray}
\hfill$\Box$\\

We recall the definition of {\em spin coherent states} \cite{P72,L73,B75}, first for a single spin:
\begin{equation}\label{D5}
| \Omega\rangle := U(\Omega)\, |s\rangle=
\sum_{m=-s}^{s}{ 2s \choose s+m}^{1/2}\,\left( \cos \frac{\theta}{2}\right)^{s+m}\,
\left( \sin \frac{\theta}{2}\right)^{s-m}\,e^{{\sf i} (s-m) \phi}\,|m\rangle
\;.
\end{equation}

The spin coherent states satisfy
\begin{equation}\label{D6}
 \left\langle \Omega \left| \left(
 \begin{array}{c}
   \op{s}_1 \\
   \op{s}_2 \\
    \op{s}_3
 \end{array} \right)
  \right|\Omega\right\rangle= s\,\Omega = s\,
 \left(
 \begin{array}{c}
   \sin\theta \,\cos\phi \\
   \sin\theta \,\sin\phi \\
   \cos\theta
 \end{array}
 \right)
 \;,
\end{equation}
see \cite{L73}, Table $1$, or \cite{FKL07}, eq.~(20). There are various relations concerning spin coherent states, see
\cite{L73} or \cite{LK15}. Here we only mention the following completeness relation
\begin{equation}\label{Did}
 {\mathbbm 1}= \frac{2s+1}{4\pi}\int_{{\mathbbm S}^2}d\Omega \left| \Omega \right\rangle \left\langle \Omega \right|
 \;,
\end{equation}
where $ {\mathbbm 1}$ denotes the identity operator in ${\mathcal H}^{(1,s)}$.
(\ref{Did}) immediately implies
\begin{equation}\label{Dtr}
 \mbox{Tr}\,A = \frac{2s+1}{4\pi}\int_{{\mathbbm S}^2}d\Omega   \left\langle \Omega \right|A \left| \Omega \right\rangle
 \;,
\end{equation}
for every linear operator $A: {\mathcal H}^{(1,s)}\rightarrow {\mathcal H}^{(1,s)}$.

Spin coherent states can be generalized to spin systems consisting of $N$ spins in a straightforward manner \cite{L73}.
The corresponding Hilbert space ${\mathcal H}^{(N,s)}$ is the $N$-fold tensor product of copies of the single spin Hilbert space
${\mathcal H}^{(1,s)}$.
The spin operators of the form ${\mathbbm 1}\otimes\ldots\otimes \op{s}_i\otimes\ldots \otimes{\mathbbm 1}$
will be denoted by $\op{s}_{\mu,i}, \;\mu=1,\ldots,N,\;i=1,2,3$ and $\op{\mathbf s}$ denotes the $N\times 3$-matrix with
these spin operators as entries. Previous definitions for the single spin case can easily be transferred, as, e.~g., $\op{s}_{\mu,\pm}$
or $\widehat{\op{s}}_{\mu,i}$, moreover
\begin{equation}\label{DU}
 U({\boldsymbol \Omega}):= U(\Omega_1)\otimes \ldots \otimes U(\Omega_N)
 \;.
\end{equation}

The common eigenbasis of the $\op{s}_{\mu,3}, \;\mu=1,\ldots,N$  will be an orthonormal basis
of ${\mathcal H}^{(N,s)}$ and its elements will be denoted by $\left| {\mathbf m}\right\rangle :=\left|m_1,\ldots,m_N\right\rangle$,
where $-s\le m_\mu \le s$ for all $\mu=1,\ldots, N$. Especially, $\left| {\mathbf s}\right\rangle := \left|s,\ldots,s\right\rangle$.

The classical $N$ spin phase space ${\mathcal S}$ is the $N$-fold cartesian product of
copies of the ${\mathbbm S}^{2}$ and its elements are denoted by ${\boldsymbol \Omega}=(\Omega_1,\ldots,\Omega_N)$.
For all ${\boldsymbol \Omega}\in{\mathcal S}$ the $N$-spin coherent state will be defined as
\begin{equation}\label{D7}
\left| {\boldsymbol \Omega}\right\rangle := U( {\boldsymbol \Omega})\left|{\mathbf s}\right\rangle
=\left| \Omega_1\right\rangle \otimes\ldots \otimes  \left| \Omega_N\right\rangle
\;.
\end{equation}

For any linear operator $A:{\mathcal H}^{(N,s)} \rightarrow {\mathcal H}^{(N,s)}$ the continuous function $g(A):{\mathcal S}\rightarrow{\mathcal S}$
defined by
\begin{equation}\label{Dcov}
g(A)({\boldsymbol\Omega}):= \left\langle {\boldsymbol\Omega} \left| A \right| {\boldsymbol\Omega}\right\rangle
\end{equation}
will be called the ``covariant symbol" of $A$, following \cite{B75}, see also Eq.~(2.15) in \cite{L73}.
The corresponding ``contravariant symbol" $G(A)$ can be defined similarly, see Eq.~(2.13) in \cite{L73} or Section \ref{sec:CS}.

As an immediate consequence of (\ref{D6}) we note
\begin{equation}\label{D8}
g \left(\op{s}_{\mu,i} \right)({\boldsymbol\Omega})= \left\langle {\boldsymbol \Omega}\left| \op{s}_{\mu,i}\right|  {\boldsymbol \Omega}\right\rangle =s\, \Omega_{\mu,i}
 \;,
\end{equation}
for all $\mu=1,\ldots,N$ and $i=1,2,3$. Here $\Omega_{\mu,i}$ denotes the $i$-th component of the $\mu$-th unit vector $\Omega_\mu$.

The completeness relation analogous to (\ref{Did}) reads:
\begin{equation}\label{DidN}
 {\mathbbm 1}= \left(\frac{2s+1}{4\pi}\right)^N \int_{{\mathcal S}}
 d{\boldsymbol\Omega} \left| {\boldsymbol\Omega}  \right\rangle \left\langle {\boldsymbol\Omega}  \right|
 \;,
\end{equation}
where $ {\mathbbm 1}$ denotes the identity operator in ${\mathcal H}^{(N,s)}$.
Also in this case (\ref{DidN}) immediately implies
\begin{equation}\label{DtrN}
 \mbox{Tr}\,A = \left(\frac{2s+1}{4\pi}\right)^N \int_{{\mathcal S}}d{\boldsymbol\Omega}
  \left\langle {\boldsymbol\Omega}\right|A \left|{\boldsymbol\Omega}\right\rangle
 \;,
\end{equation}
for every linear operator $A: {\mathcal H}^{(N,s)}\rightarrow {\mathcal H}^{(N,s)}$.

We recall some well-known definitions:
For any vector ${\boldsymbol\phi}\in{\mathcal H}^{(N,s)}$  its vector norm will be denoted by
$\|{\boldsymbol\phi}\|:=\sqrt{\langle {\boldsymbol\phi}| {\boldsymbol\phi}\rangle}$.
For any linear operator $A:{\mathcal H}^{(N,s)} \rightarrow {\mathcal H}^{(N,s)}$ the corresponding ``operator norm" is
defined by $\|A \| := \sup_{\|{\boldsymbol\phi}\|=1}\| A {\boldsymbol\phi} \|$. If $A$ is Hermitean (or, more general, normal)
its operator norm equals its ``spectral radius" $\rho(A):= \max \{ |\lambda_1|,\ldots |\lambda_n|\}$, where $\lambda_1,\ldots \lambda_n$
are the eigenvalues of $A$. Hence, in this case,
$\|A \|=\sup_{\|{\boldsymbol\phi}\|=1} \left| \left\langle {\boldsymbol\phi} \left|A\right| {\boldsymbol\phi}\right\rangle \right|.$

The unit sphere ${\mathbbm S}^2$ is a compact metric space, actually a Riemannian manifold.
The same holds for the finite cartesian product ${\mathcal S}={\mathbbm S}^2\times \ldots \times {\mathbbm S}^2$ of its copies.
The linear space of real continuous functions $f:{\mathcal S} \rightarrow {\mathcal S}$ will be denoted by ${{\mathcal C}}({\mathcal S})$.
It is complete w.~r.~t.~the so-called sup-norm $\| f\|:= \sup_{x\in{\mathcal S}}|f(x)|$ since ${\mathcal S}$ is compact.\\

For later purposes we will show the following
\begin{lemma}\label{LemmaSup}
Let $n\mapsto A_n$ be a sequence of Hermitean operators in ${\mathcal H}^{(N,s)}$ converging to an operator $A$ w.~r.~t.~the operator norm,
then the sequence $n\mapsto g(A_n)$ will converge to
$g(A)$ w.~r.~t.~the sup-norm in ${\mathcal C}({\mathcal S})$.
\end{lemma}
{\bf Proof}: By assumption, for each $\epsilon>0$ there exists an $n_0\in{\mathbbm N}$ such that for all $n \ge n_0$
and all ${\boldsymbol\phi}\in {\mathcal H}^{(N,s)}$ satisfying $\| {\boldsymbol\phi}\|=1$ it holds that
$\left| \left\langle {\boldsymbol\phi} \left| A_n -A\right| {\boldsymbol\phi}\right\rangle \right| < \epsilon$.
This statement also applies to the subset of spin coherent states ${\boldsymbol\Omega}$ and hence
$\sup_{\boldsymbol\Omega} \left| \left\langle {\boldsymbol\Omega} \left| A_n -A\right| {\boldsymbol\Omega}\right\rangle \right| < \epsilon$,
which immediately implies the desired convergence of the covariant symbols w.~r.~t.~the sup-norm.         \hfill$\Box$\\

\section{Classical limit of Gibbs states}\label{sec:G}

Gibbs states are statistical operators $G^{(N,s)}$ living in the Hilbert spaces ${\mathcal H}^{(N,s)}$
and hence cannot be related in a direct manner.
As explained in the Introduction, we will follow the usual strategy to rather consider the covariant symbols
$g(G^{(N,s)})=\left\langle {\boldsymbol\Omega}\left|G^{(N,s)}\right| {\boldsymbol\Omega}\right\rangle $ that are continuous functions
on the same classical phase space ${\mathcal S}$. In our case we want to prove that the covariant symbols of (suitably scaled)  Gibbs states
converge for $s\rightarrow\infty$ to the classical Gibbs state $G^{\text{(cl)}}$ w.~r.~t.~the sup-norm in
${\mathcal C}({\mathcal S})$. In a first step, we will ignore the trace (partition function)
and only consider the classical limit of $\exp(-\beta \widehat{H}^{(s)})$.

\subsection{Classical limit of $\exp(-\beta \widehat{H}^{(s)})$}\label{sec:GE}

Our aim is to prove that the classical limit of $W^{(N,s)}:=\exp(-\beta \widehat{H}^{(s)}))$ will be the function $W^{\text{(cl)}}$ on ${\mathcal S}$
given by $W^{\text{(cl)}}=\exp(-\beta {H}^{\text{(cl)}})$.
There exist some useful related results about covariant symbols of monomials in the spin operators, see \cite{L73} and \cite{FKL07},
hence our strategy will be to approximate $W^{(N,s)}=\exp(-\beta \widehat{H}^{(s)}))$
by polynomials (truncated Taylor series) $W^{(N,s,n)}$ such that $\lim_{n\rightarrow\infty} W^{(N,s,n)}=W^{(N,s)}$
and hence $\lim_{n\rightarrow\infty} g\left((W^{(N,s,n)}\right)=g\left(W^{(N,s)}\right)$ by virtue of Lemma \ref{LemmaSup}.
If we could show that each sequence $s\mapsto g\left(W^{(N,s,n)}\right)$ has a classical
limit $W^{(N,n,\text{(cl)})}$ for $s\rightarrow\infty$,
then the limit of $W^{(N,n,\text{(cl)})}$ for $n\rightarrow\infty$ would be a candidate for $W^{\text{(cl)}}$.
However, here we encounter the problem that, in general, the iterated limits of a double sequence need not coincide.

In this context we cite the following theorem, see, e.~g., \cite{O59} \S 1014,
\begin{theorem}(Moore-Osgood)\label{TMO}
 Let $\{u_{mn}\}$ be a double sequence with values in a complete metric space. If
 \begin{equation}\label{TMO1}
  \lim_{n\rightarrow\infty}u_{mn}=q_m \quad \mbox{exists for each } m
 \end{equation}
and if
\begin{equation}\label{TMO2}
  \lim_{m\rightarrow\infty}u_{mn}=p_n \quad \mbox{ (uniformly in } n \mbox{) likewise exists,}
 \end{equation}
then the double limit and the two iterated limits of $\{u_{mn}\}$ exist and
 \begin{equation}\label{TMO3}
  \lim_{ n\rightarrow\infty \atop m\rightarrow\infty}u_{mn}=
  \lim_{n\rightarrow\infty}\lim_{m\rightarrow\infty}u_{mn}=
  \lim_{m\rightarrow\infty}\lim_{n\rightarrow\infty}u_{mn}
  \;.
 \end{equation}
\end{theorem}

In view of this theorem we must establish the existence of the two limits for $n\rightarrow\infty$ and  $s\rightarrow\infty$
such that one of these limits is assumed in a uniform manner. It turns out that, although the limit $n\rightarrow\infty$,
i.~e., the convergence of the exponential series, is well-known, it is not uniform in $s$, since the speed of convergence
depends in a crucial way on the spectrum of the Hamiltonian.
Therefore we will try the second possibility by showing that the classical limit $s\rightarrow\infty$  of the
covariant symbols of the polynomials $W^{(N,s,n)}$ exists and is uniform in $n$.

To formulate the first result it will be convenient to merge the indices $\mu,i$ into a multi-index $j=(\mu,i)$.
Recall that $\widehat{\op{s}}_j=\widehat{\op{s}}_{\mu,i}= \frac{1}{s}\,\op{s}_{\mu,i}$. We consider the covariant
symbol of finite products of these operators and obtain the following result that is analogous to Lemma $4$
appearing in \cite{FKL07} without a proof:
\begin{lemma}\label{LemmaProduct}
\begin{equation}\label{G1}
  \left| \left\langle {\boldsymbol \Omega} \left| \widehat{\op{s}}_{j_1}\ldots \widehat{\op{s}}_{j_p}\right|  {\boldsymbol \Omega} \right\rangle
  - \Omega_{j_1}\ldots \Omega_{j_p}\right| \le \frac{p}{\sqrt{2 s}}
  \;.
\end{equation}
\end{lemma}
{\bf Proof}: By induction over $p$.\\
(i) For $p=1$ it follows by (\ref{D8}) that
\begin{equation}\label{G2}
 \left| \left\langle {\boldsymbol \Omega} \left| \widehat{\op{s}}_{j_1}\right|  {\boldsymbol \Omega} \right\rangle
  - \Omega_{j_1}\right| =0
  \;,
\end{equation}
and hence (\ref{G1}) is satisfied.\\
(ii) Next we assume that (\ref{G1}) holds for some $p$ and will prove its validity for $p+1$. To this end consider
\begin{eqnarray}
\label{G3a}
  \left\langle {\boldsymbol \Omega} \left| \widehat{\op{s}}_{j_1}\ldots \widehat{\op{s}}_{j_{p+1}}\right|  {\boldsymbol \Omega} \right\rangle
   &\stackrel{(\ref{D7})}{=}&
   \left\langle \mathbf{s} \left|U^\ast({\boldsymbol \Omega})\, \widehat{\op{s}}_{j_1}\ldots \widehat{\op{s}}_{j_p}\,
  U({\boldsymbol \Omega})\,U^\ast({\boldsymbol \Omega})\,\widehat{\op{s}}_{j_{p+1}}\,U({\boldsymbol \Omega})
   \right|   \mathbf{s} \right\rangle\\
   \label{G3b}
   &=&
   \sum_{\left|{\mathbf m}\right\rangle}
    \left\langle \mathbf{s} \left|U^\ast({\boldsymbol \Omega})\, \widehat{\op{s}}_{j_1}\ldots \widehat{\op{s}}_{j_p}\,U({\boldsymbol \Omega})
    \right| {\mathbf m}\right\rangle
    \left\langle{\mathbf m}\left|U^\ast({\boldsymbol \Omega})\, \widehat{\op{s}}_{j_{p+1}}\, U({\boldsymbol \Omega})\right| {\mathbf s}\right\rangle\\
    \nonumber
    &=&
    \left\langle \mathbf{s} \left|U^\ast({\boldsymbol \Omega})\, \widehat{\op{s}}_{j_1}\ldots \widehat{\op{s}}_{j_p}\,U({\boldsymbol \Omega})
    \right| {\mathbf s}\right\rangle
    \left\langle{\mathbf s}\left|\widehat{\op{s}}_{j_{p+1}}({\boldsymbol \Omega})\right| {\mathbf s}\right\rangle\\
    \label{G3c}
    &&+ \sum_{\left|{\mathbf m}\right\rangle\neq\left|{\mathbf s}\right\rangle}
    \left\langle \mathbf{s} \left|U^\ast({\boldsymbol \Omega})\, \widehat{\op{s}}_{j_1}\ldots \widehat{\op{s}}_{j_p}\,U({\boldsymbol \Omega})
    \right| {\mathbf m}\right\rangle
    \left\langle{\mathbf m}\left|\widehat{\op{s}}_{j_{p+1}}({\boldsymbol \Omega})\right| {\mathbf s}\right\rangle\\
    \label{G3d}
    &=:& T_1+T_2
    \;.
\end{eqnarray}
The first term $T_1$ of (\ref{G3c}) can be written as
\begin{eqnarray}
\label{G4a}
  T_1 &\stackrel{(\ref{D8})}{=}&  \left\langle \mathbf{s} \left|U^\ast({\boldsymbol \Omega})\,
  \widehat{\op{s}}_{j_1}\ldots \widehat{\op{s}}_{j_p}\,U({\boldsymbol \Omega})\right| {\mathbf s}\right\rangle
    \; \Omega_{j_{p+1}} \\
    \label{G4b}
    &=& \Omega_{j_1}\ldots \Omega_{j_p}\Omega_{j_{p+1}}+
    \left(\left\langle {\boldsymbol \Omega} \left|  \widehat{\op{s}}_{j_1}\ldots \widehat{\op{s}}_{j_p}\right| {\boldsymbol \Omega}\right\rangle
    -\Omega_{j_1}\ldots \Omega_{j_p}
    \right)\Omega_{j_{p+1}}\;,
 \end{eqnarray}
and hence
\begin{eqnarray}
\label{G5a}
  \left|T_1-  \Omega_{j_1}\ldots \Omega_{j_p}\Omega_{j_{p+1}} \right|  &\stackrel{(\ref{G4b})}{=}&
  \left|
   \left(\left\langle {\boldsymbol \Omega} \left|  \widehat{\op{s}}_{j_1}\ldots \widehat{\op{s}}_{j_p}\right| {\boldsymbol \Omega}\right\rangle
    -\Omega_{j_1}\ldots \Omega_{j_p}
    \right)\Omega_{j_{p+1}}
    \right|\\
    \label{G5b}
    &=&
    \left|
   \left\langle {\boldsymbol \Omega} \left|  \widehat{\op{s}}_{j_1}\ldots \widehat{\op{s}}_{j_p}\right| {\boldsymbol \Omega}\right\rangle
    -\Omega_{j_1}\ldots \Omega_{j_p}\right|
    \left|\Omega_{j_{p+1}}\right|\\
    \label{G5c}
    &\le&
     \left|
   \left\langle {\boldsymbol \Omega} \left|  \widehat{\op{s}}_{j_1}\ldots \widehat{\op{s}}_{j_p}\right| {\boldsymbol \Omega}\right\rangle
    -\Omega_{j_1}\ldots \Omega_{j_p}\right|\\
    \label{G5d}
    &\le& \frac{p}{\sqrt{2s}}
    \;.
\end{eqnarray}
In (\ref{G5c}) we have used the property $ \left|\Omega_{j_{p+1}}\right|\le 1$ of the components of a unit vector and
in (\ref{G5d}) the induction hypothesis (\ref{G1}).

For the second term $T_2$ of (\ref{G3c}) we let $j_{p+1}=(\mu,i)$ and note that the second factor
\begin{equation}\label{G6}
 \left\langle{\mathbf m}\left|\widehat{\op{s}}_{j_{p+1}}({\boldsymbol \Omega})\right| {\mathbf s}\right\rangle
 =
 \left\langle{\mathbf m}\left|\widehat{\op{s}}_{\mu,i}({\boldsymbol \Omega})\right| {\mathbf s}\right\rangle
 \end{equation}
vanishes unless $m_\mu=s-1$ and $m_\nu=s$ for all $\nu\neq \mu$. This follows from the tridiagonal form of
$\widehat{\op{s}}_{\mu,i}({\boldsymbol \Omega})$ due to (\ref{D4a}) - (\ref{D4c}). Hence the sum for $T_2$
in (\ref{G3c}) coalesces into a single term with second factor
\begin{equation}\label{G7}
  \left\langle{\mathbf m}\left|\widehat{\op{s}}_{j_{p+1}}({\boldsymbol \Omega})\right| {\mathbf s}\right\rangle
 =\left\langle s-1 \left| \widehat{\op{s}}_i\left(\Omega_\mu\right) \right| s \right\rangle
 =\frac{1}{s}\sqrt{2s}\,{s}_{i,-}\left(\Omega_\mu\right)
 \;,
\end{equation}
where we have used (\ref{D3}) and (\ref{Dspin}) for $m=s-1$. Its absolute value can be estimated according to
\begin{equation}\label{G8}
 \left| \left\langle{\mathbf m}\left|\widehat{\op{s}}_{j_{p+1}}({\boldsymbol \Omega})\right| {\mathbf s}\right\rangle\right|
 =\left|\frac{1}{s}\sqrt{2s}\,\op{s}_{i,-}\left(\Omega_\mu\right) \right|
\stackrel{(\ref{DC1})}{\le} \frac{1}{s}\sqrt{2s}\, {\scriptsize\frac{1}{2}}= \frac{1}{\sqrt{2 s}}
\;.
\end{equation}
The absolute value of the first factor of $T_2$ in  (\ref{G3c}) can be estimated by
\begin{equation}\label{G9}
 \left|
 \left\langle \mathbf{s} \left|U^\ast({\boldsymbol \Omega})\, \widehat{\op{s}}_{j_1}\ldots \widehat{\op{s}}_{j_p}\,U({\boldsymbol \Omega})
 \right| {\mathbf m}\right\rangle
 \right|
 \le
 \| \mathbf{s} \| \|\widehat{\op{s}}_{j_1} \| \ldots \|\widehat{\op{s}}_{j_p} \|  \| \mathbf{m} \|
 =1
 \;,
\end{equation}
since the operator norm $\|\widehat{\op{s}}_{j_\ell} \|$, i.~e., the largest absolute value of the eigenvalues of
$\widehat{\op{s}}_{j_\ell}$, is unity for $\ell=1,\ldots,p$ due to the scaling. Summarizing,
\begin{equation}\label{G10}
  \left| T_2 \right| \stackrel{(\ref{G8},\ref{G9})}{\le}\frac{1}{\sqrt{2 s}}
  \;,
\end{equation}
and hence
\begin{equation}\label{G11}
   \left| T_1+T_2-  \Omega_{j_1}\ldots \Omega_{j_p}\Omega_{j_{p+1}}  \right|
   \le \left| T_1-  \Omega_{j_1}\ldots \Omega_{j_p}\Omega_{j_{p+1}}  \right| +\left| T_2\right|
   \stackrel{(\ref{G5d},\ref{G10})}{\le}
   \frac{p}{\sqrt{2s}}+\frac{1}{\sqrt{2 s}} =\frac{p+1}{\sqrt{2s}}
   \;,
\end{equation}
thereby confirming the induction claim (\ref{G1}) for $p+1$. \hfill$\Box$\\

Lemma \ref{LemmaProduct} means that the covariant symbols of monomials (and hence of polynomials) in the spin operators converge
point-wise to their classical counterparts w.~r.~t.~the sup-norm in ${\mathcal C}\left( {\mathcal S} \right)$. But due to
factor $p$ at the r.~h.~s.~of  (\ref{G1}) this convergence is not uniform in the set of all monomials. We will see, however, that uniform
convergence can be accomplished for the sequence of polynomials that converge to the (non-normalized) Gibbs state for a given Hamiltonian.

To this end we consider a scaled Hamiltonian $\widehat{H}^{(s)}$ of the form
\begin{equation}\label{G11}
 \widehat{H}^{(s)}= \sum_{q=1}^{p}\sum_{j_1,\ldots,j_q} J_{j_1,\ldots,j_q}\,\widehat{\op{s}}_{j_1}\ldots \widehat{\op{s}}_{j_q}
 \;.
\end{equation}
Here we have again adopted the multi-index notation $j=(\mu,i)$. We have thus assumed that, generally, the Hamiltonian
is an inhomogenous polynomial in the spin operators of maximal degree $p$.
The second sum in (\ref{G11}) runs over a subset of all $(3\,N)^q$ possible values;
the total number of monomials that appear in the multiple sum of (\ref{G11}) will be denoted by $L$.
The real coupling coefficients  $J_{j_1,\ldots,j_q}$ are bounded in absolute value by
\begin{equation}\label{G12}
J:=\mbox{Max}_{q}\,\mbox{Max}_{j_1,\ldots,j_q} \left|J_{j_1,\ldots,j_q} \right|
\;.
\end{equation}
For example, the Heisenberg coupling scheme $\widehat{H}^{(s)}=\sum_{1\le\mu<\nu\le N}J_{\mu\nu}\, \widehat{\op{\mathbf s}}_\mu\cdot \widehat{\op{\mathbf s}}_\nu$
satisfies $p=2$ and the sum runs over $L=3 {N \choose 2}$ possible values if the scalar products are expanded.
Due to its dependence on $s$ the expression (\ref{G11}) should rather be regarded as a family of Hamiltonian operators acting on
Hilbert spaces ${\mathcal H}^{(N,s)}$, but we will nevertheless refer to (\ref{G11}) as ``the" Hamiltonian without danger of confusion.

The classical Hamiltonian corresponding to (\ref{G11}) will be defined as
\begin{equation}\label{G13}
  H^{\text{(cl)}}:=\sum_{q=1}^{p} \sum_{j_1,\ldots,j_q} J_{j_1,\ldots,j_q}\,\Omega_{j_1}\ldots \Omega_{j_q}
  \;,
\end{equation}
and hence as a polynomial in the variables $\Omega_{j_\ell}$ satisfying $ H^{\text{(cl)}}\in{\mathcal C}\left( {\mathcal S} \right)$.

Next we consider a parameter $\beta>0$ with the physical meaning of a (dimensionless) inverse temperature and the corresponding
non-normalized Gibbbs state
\begin{equation}\label{G14}
 W^{(s)}:=\exp\left( - \beta \widehat{H}^{(s)}\right) = \lim_{n\rightarrow\infty} W_n^{(s)}
 \;,
\end{equation}
where
\begin{equation}\label{G15}
  W_n^{(s)}:=\sum_{k=0}^{n} W_{n,k}^{(s)}:=\sum_{k=0}^{n} \frac{(-\beta)^k}{k!}\,\widehat{H}^{(s)k}
  \;.
\end{equation}
The exponential series (\ref{G14}) converges in any Hilbert space ${\mathcal H}^{(N,s)}$ w.~r.~t.~all
equivalent operator topologies, especially in the operator norm. This implies that the covariant symbols
$\left\langle {\boldsymbol \Omega}\left| W_n^{(s)} \right|  {\boldsymbol \Omega}\right\rangle$ converge
to $\left\langle {\boldsymbol \Omega}\left| \exp\left( - \beta \widehat{H}^{(s)}\right) \right|  {\boldsymbol \Omega}\right\rangle$
in the sup-norm, see Lemma \ref{LemmaSup}.

The classical counterparts to (\ref{G14}) and (\ref{G15}) are
\begin{eqnarray}
\label{G15a}
  W_{n,k}^{\text{cl}}&:=& \frac{(-\beta)^k}{k!}\,\left( H^{\text{(cl)}}\right)^k, \\
  \label{G15b}
  W_{n}^{\text{cl}}&:=&\sum_{k=0}^{n}\frac{(-\beta)^k}{k!}\,\left( H^{\text{(cl)}}\right)^k, \\
  \label{G15c}
   W^{\text{cl}}&:=&\exp\left( - \beta H^{\text{(cl)}}\right) = \lim_{n\rightarrow\infty} W_n^{\text{(cl)}}
 \;.
\end{eqnarray}
In the classical case the exponential series (\ref{G15c}) converges in the sup-norm to a continuous function
 $W^{\text{cl}}\in {\mathcal C}\left( {\mathcal S} \right)$.

It is clear that the $k$-th power of $\widehat{H}^{(s)}$ is again a polynomial in the spin operators, and hence of the form
 \begin{equation}\label{G16}
  \widehat{H}^{(s)k} = \sum_{\ell_1,\ldots,\ell_r}K_{\ell_1,\ldots,\ell_r}\,\widehat{\op{s}}_{\ell_1}\ldots \widehat{\op{s}}_{\ell_r}
  \;.
 \end{equation}
 We will not dwell upon the well-known details on how to calculate the coefficients of this polynomial, but only use the following facts:
 The total number of terms in the sum of (\ref{G16}) is $L^k$, and, further,
 \begin{equation}\label{G17}
   r\le p\,k, \mbox{ and } \left|K_{\ell_1,\ldots,\ell_r}\right| \le J^k \mbox{ for all } \ell_1,\ldots,\ell_r
   \;.
 \end{equation}
 Here we have implicitly assumed that $ \widehat{H}^{(s)k}$ is obtained by a termwise $k$-fold multiplication of the polynomial
 $ \widehat{H}^{(s)}$ without further collecting terms. The classical counterpart of $ \widehat{H}^{(s)k}$ can be written as
 \begin{equation}\label{G18}
   H^{\text{(cl)}k}=
  \sum_{\ell_1,\ldots,\ell_r}K_{\ell_1,\ldots,\ell_r}\,\Omega_{\ell_1}\ldots \Omega_{\ell_r}
  \;.
 \end{equation}
 We conclude
 \begin{eqnarray}
\label{G19a}
   \left|\left\langle {\boldsymbol\Omega}\left| W_{n,k}^{(s)}\right| {\boldsymbol\Omega}\right\rangle-W_{n,k}^{\text{(cl)}} \right| &=&
   \left|
   \frac{(-\beta)^k}{k!}\sum_{\ell_1,\ldots,\ell_r}K_{\ell_1,\ldots,\ell_r}\left(
   \left\langle {\boldsymbol\Omega}\left|\widehat{\op{s}}_{\ell_1}\ldots \widehat{\op{s}}_{\ell_r} \right| {\boldsymbol\Omega}\right\rangle
   - \Omega_{\ell_1}\ldots \Omega_{\ell_r}
   \right)
   \right| \\
   \label{G19b}
   &\le&
    \frac{\beta^k}{k!}\sum_{\ell_1,\ldots,\ell_r}\left| K_{\ell_1,\ldots,\ell_r}\right|\,
    \left|\left\langle{\boldsymbol\Omega}\left|\widehat{\op{s}}_{\ell_1}\ldots \widehat{\op{s}}_{\ell_r} \right| {\boldsymbol\Omega}\right\rangle
   - \Omega_{\ell_1}\ldots \Omega_{\ell_r} \right|\\
   \label{G19c}
   &\stackrel{(\ref{G1},\ref{G17})}{\le}&
    \frac{\beta^k}{k!}\,L^k\,J^k\,\frac{p\,k}{\sqrt{2s}}
    \;,
\end{eqnarray}
and further
\begin{eqnarray}
\label{G20a}
   \left|\left\langle {\boldsymbol\Omega}\left| W_{n}^{(s)}\right| {\boldsymbol\Omega}\right\rangle-W_{n}^{\text{(cl)}} \right| &\le&
   \sum_{k=1}^{n} \left|\left\langle {\boldsymbol\Omega}\left| W_{n,k}^{(s)}\right| {\boldsymbol\Omega}\right\rangle-W_{n,k}^{\text{(cl)}} \right|
    \\
    \label{G20b}
   &\stackrel{(\ref{G19c})}{\le}&  \sum_{k=1}^{n} \frac{\beta^k}{k!}\,L^k\,J^k\,\frac{p\,k}{\sqrt{2s}}    \\
   \label{G20c}
   &=& \frac{\beta L J p}{\sqrt{2s}} \sum_{k=1}^{n} \frac{\beta^{k-1}}{(k-1)!}\,L^{k-1}\,J^{k-1}    \\
   \label{G20d}
   &\le& \frac{\beta L J p}{\sqrt{2s}} e^{\beta\,L\,J}
   \;.
\end{eqnarray}
In (\ref{G20a}) we have omitted the term corresponding to $k=0$ since
$\left\langle {\boldsymbol\Omega}\left| W_{n,0}^{(s)}\right| {\boldsymbol\Omega}\right\rangle = W_{n,0}^{\text{(cl)}}=1$.
Note that the bound (\ref{G20d}) is independent of $n$ and can be made arbitrarily small for $s\rightarrow\infty$.

In view of the considerations at the beginning of this subsection we have shown that both limits
$n\rightarrow\infty$ and  $s\rightarrow\infty$ of $g\left(W^{(N,s,n)}\right)$ exist and that the limit
$s\rightarrow\infty$ is assumed in a uniform manner. Hence, by virtue of the Theorem \ref{TMO} (Moore-Osgood)
the double limit $\lim_{n\rightarrow\infty \atop s\rightarrow\infty}g\left(W^{(N,s,n)}\right)$ exists
and is equal to the two iterated limits
\begin{equation}\label{G21}
  \lim_{n\rightarrow\infty}\, \lim_{s\rightarrow\infty}\,g\left(W^{(N,s,n)}\right)
  =\lim_{n\rightarrow\infty}\,W_n^{\text{(cl)}}=W^{\text{(cl)}}=\exp\left( - \beta H^{\text{(cl)}}\right)
  \;,
\end{equation}
and
\begin{equation}\label{G22}
  \lim_{s\rightarrow\infty}\, \lim_{n\rightarrow\infty}\,g\left(W^{(N,s,n)}\right)
  =\lim_{s\rightarrow\infty}\, g\left(W^{(N,s)} \right)=W^{\text{(cl)}}
  \;.
\end{equation}
Summarizing, we have proven the following
\begin{theorem}\label{TGE}
With the preceding definitions,
\begin{equation}\label{G23}
  \lim_{s\rightarrow\infty}\,g\left(\exp(-\beta \widehat{H}^{(s)}))\right)=\exp(-\beta {H}^{\text{(cl)}})\,
 \end{equation}
where the limit is understood w.~r.~t.~the sup-norm in ${\mathcal C}\left( {\mathcal S}\right)$.
\end{theorem}

\subsection{Classical limit of Gibbs states}\label{sec:GG}

Next we consider the trace of $\exp(-\beta \widehat{H}^{(s)})= W^{(N,s)}$. By virtue of (\ref{Dcov}) and  (\ref{DtrN}) we have
\begin{equation}\label{GG1}
  \mbox{Tr }\left(  W^{(N,s)} \right)=\left( \frac{2s+1}{4\pi}\right)^N \int_{\mathcal S}d{\boldsymbol\Omega}\, g(W^{(N,s)})
  \;.
\end{equation}

According to Theorem \ref{TGE} the covariant symbol $g(W^{(N,s)})$ converges  to  $\exp(-\beta {H}^{\text{(cl)}}))$ for
$s\rightarrow\infty$ in the sup-norm, hence the integral  $\int_{\mathcal S}d{\boldsymbol\Omega}\, g(W^{(N,s)})$ converges to
$\int_{\mathcal S}d{\boldsymbol\Omega}\,\exp(-\beta {H}^{\text{(cl)}}))$.
This implies convergence of the scaled trace according to (\ref{GG1}) to the classical partition function
\begin{equation}\label{GG2}
 \lim_{s\rightarrow\infty}\left(2s+1\right)^{-N}  \mbox{Tr }\left(  W^{(N,s)} \right)
 =
 \left(4\pi \right)^{-N} \int_{\mathcal S}d{\boldsymbol\Omega}\, \exp(-\beta {H}^{\text{(cl)}})):= Z^{\text{(cl)}}(\beta)
  \;.
\end{equation}
For an alternative derivation of (\ref{GG2}) see \cite{L73}.

Recall the usual definition of the Gibbs state $G^{(N,s)}$:
\begin{equation}\label{GG3}
 G^{(N,s)}:=\left( \mbox{Tr}\,\exp(-\beta \widehat{H}^{(s)})\right)^{-1} \,\exp(-\beta \widehat{H}^{(s)})
 =
 \left( \mbox{Tr}\, W^{(N,s)}\right)^{-1} \, W^{(N,s)}
 \;.
\end{equation}
This means that our main result about the classical limit of Gibbs states has to be construed as the following statement about the limit
of the covariant symbol of the suitably scaled quantum Gibbs state:
\begin{theorem}\label{TGG}
 Under the preceding definitions the following holds:
 \begin{equation}\label{GG4}
 \lim_{s\rightarrow\infty}\left(2s+1\right)^{N} g\left(  G^{(N,s)}\right)= \left( Z^{\text{(cl)}}(\beta)\right)^{-1}\,\exp(-\beta {H}^{\text{(cl)}}))
 =: G^{\text{(cl)}}
 \;,
  \end{equation}
  where the limit is be understood w.~r.~t.~the sup-norm on ${\mathcal C}\left( {\mathcal S}\right)$.
\end{theorem}
The {\bf proof} follows immediately from (\ref{G23}) and (\ref{GG2}). \hfill$\Box$\\

We stress that a twofold scaling is involved: First, we consider the scaled Hamiltonian $\widehat{H}^{(s)}$
where all spin operators $\op{s}_i$ are replaced by $\widehat{\op{s}}_i = \frac{1}{s}\op{s}_i$.
Second, the Gibbs state $G^{(N,s)}$  w.~r.~t.~$\widehat{H}^{(s)}$ has to be scaled with a factor $\left(2s+1\right)^{N}$
in order to obtain a sensible classical limit for $s\rightarrow\infty$.

\subsection{Example: The Heisenberg dimer}
\label{sec:HD}

As a solvable example illustrating the results of this section we consider the spin dimer ($N=2$) with Hamiltonian
\begin{equation}\label{HD1}
  \widehat{H}^{(s)}=\widehat{\op{s}}_1\cdot \widehat{\op{s}}_2
  \;,
\end{equation}
and arbitrary spin quantum number $s$. As it is well-known, the eigenvalues of (\ref{HD1}) are of the form
\begin{equation}\label{HD2}
 E_S=\frac{1}{2 s^2}\left(S(S+1)-2 s(s+1)\right)
 \;,
\end{equation}
with $(2S+1)$-fold degeneracy
for $S=0,1,\ldots,2s$, and the corresponding eigenvectors are
\begin{equation}\label{HD3}
 \left| S,m\right\rangle := \sum CG(s,m_1;s,m_2;S,m)\, \left| m_1,m_2\right\rangle
\end{equation}
for $m=-S,\ldots,S$, where $CG(s,m_1;s,m_2;S,m)$ denote the Clebsch-Gordan coefficients.

\begin{figure}[t]
\centering
\includegraphics[width=1.0\linewidth]{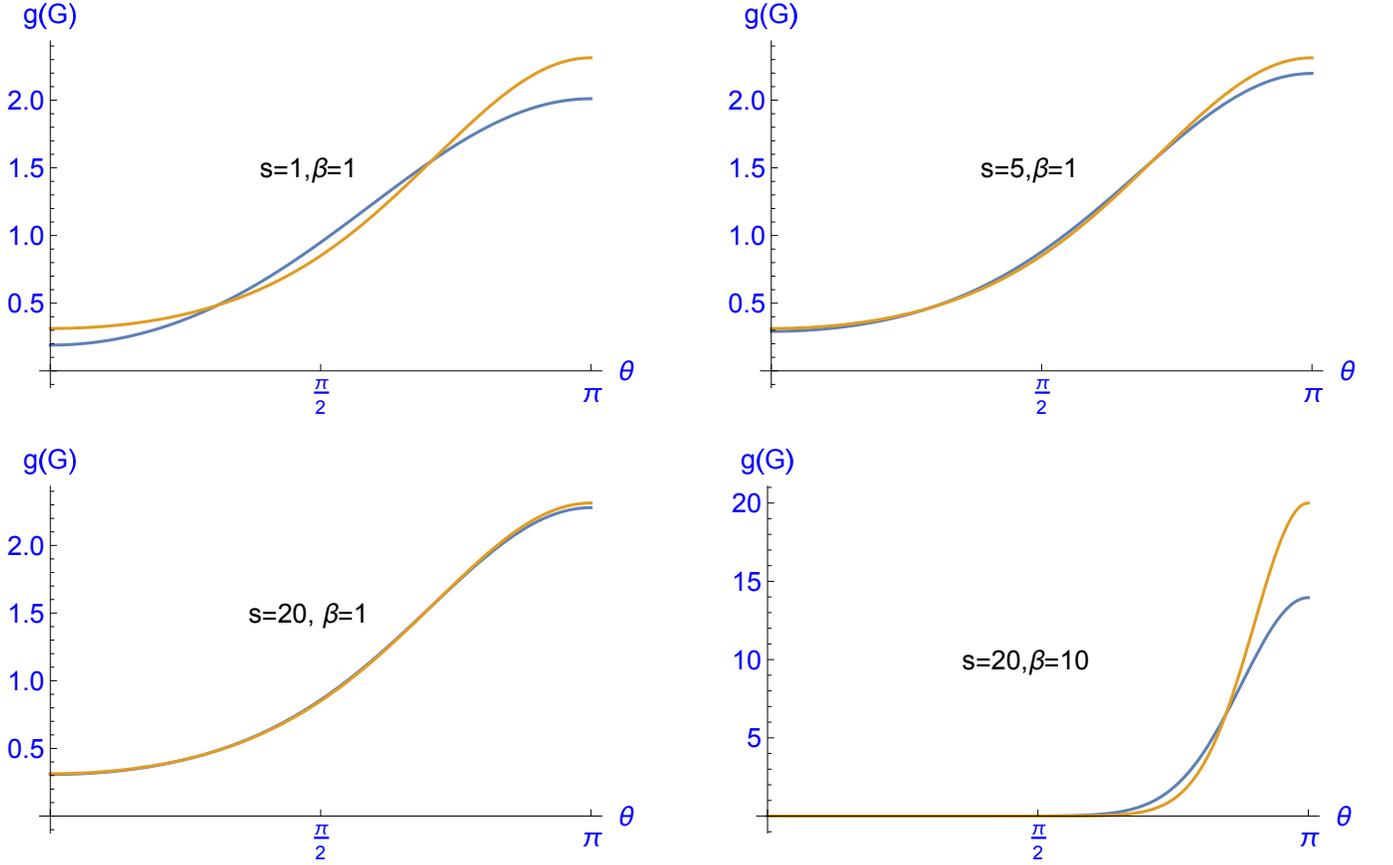}
	\caption{The covariant symbol $g(G)$ of the scaled Gibbs states of the Heisenberg dimer according to (\ref{HD4})
   as a function of $\theta=\Omega_1\cdot \Omega_2$ for different values of
  inverse temperature $\beta$ and spin quantum numbers $s$ (blue curves). The classical limit (\ref{HD6}) is also shown (dark yellow curves).
  For $\beta=1$ we observe an increasing convergence of the quantum Gibbs states to the classical one for $s=1,5,20$.
  On the other hand, convergence is significantly slower at $\beta=10$ and $\theta\approx \pi$.}
\label{FIGG}
\end{figure}

By using explicit representations of the Clebsch-Gordan coefficients \cite{AS65} and spin-coherent states (\ref{D5})
we obtain after some calculations a closed form of the covariant symbol of the scaled Gibbs states:
\begin{eqnarray}
 \nonumber
  g(G) &=&\frac{(2 s+1)^2}{Z} \sum _{S=0}^{2 s} \exp \left(-\frac{\beta  \left(\frac{1}{2} S
   (S+1)-s (s+1)\right)}{s^2}\right)\frac{ (1+2 S) \sin ^{2 s}(\theta ) (2 s)! (2s)!}{2^{2 s} (2 s-S)! (1+2 s+S)!}\times\\
   \label{HD4}
   &&
  \sum _{m=s-S}^s \frac{(-m+s+S)!
   \tan ^{2 m}\left(\frac{\theta}{2}\right)}{(-m+s)! (m-s+S)! (s-m)!}
   \;,
\end{eqnarray}
where $Z$ denotes the  partition function
\begin{equation}\label{HD5}
Z= \sum _{S=0}^{2 s} (2 S+1) \exp \left(-\frac{\beta}{s^2}
 \left(\frac{1}{2} S (S+1)-s(s+1)\right)\right)
\;.
\end{equation}
Here we have restricted the dependence on $(\Omega_1,\Omega_2)$ to the angle $\theta$ between these vectors,
i.~e., $\cos\theta = \Omega_1\cdot \Omega_2$. This considerably simplifies the expression (\ref{HD4})
and contains all relevant information due to rotational symmetry.
We have plotted the results for the values $\beta=1$ and $s=1,5,20$ together with the classical Gibbs state
\begin{equation}\label{HD6}
  G^{\text{(cl)}}=\frac{e^{-\beta  \cos (\theta )} \beta }{\sinh (\beta )}
  \;,
\end{equation}
see Figure \ref{FIGG}.
For $\beta=1$ a clear convergence of $g(G)$ to $G^{\text{(cl)}}$ can be observed. However,
the velocity of convergence slows down for larger values of $\beta$,
as can be seen in Figure \ref{FIGG} for the value $\beta=10$. Note that for $\beta\rightarrow\infty$ the
quantum Gibbs states approach the projector onto the (unique) ground state corresponding to $S=0$ and
$G^{\text{(cl)}}$ approaches the singular measure concentrated on $\theta=\pi$.
Hence a uniform convergence w.~r.~t.~$\beta$ in the sup-norm is ruled out
which explains the present finding.

\section{Explicit calculation of covariant symbols of monomials}
\label{sec:A}

\subsection{General part}
\label{sec:AA}

We work in the Hilbert space ${\mathcal H}^{(s)}\cong {\mathbbm C}^{2s+1}$ of a single spin and
consider monomials of scaled spin operators of the form
\begin{equation}\label{A1}
 {\sf M}=\widehat{\op{s}}_{i_1}\ldots \widehat{\op{s}}_{i_n}\;,
\end{equation}
for arbitrary $n=1,2,3,\ldots$. The  considerations used in the proof of Lemma \ref{LemmaProduct}
lead us to a method of explicitly calculating the covariant symbol $g({\sf M})$. It can be written as
\begin{eqnarray}
\label{A2a}
 g({\sf M}) &=&\left\langle \Omega \left|  \widehat{\op{s}}_{i_1}\ldots \widehat{\op{s}}_{i_n}\right| \Omega \right\rangle \\
 \label{A2b}
 &=& \left\langle s \left| U^\ast(\Omega)\, \widehat{\op{s}}_{i_1}\ldots \widehat{\op{s}}_{i_n}U(\Omega)\right| s \right\rangle\\
 \label{A2c}
 &=& \left\langle s \left| \widetilde{\op{s}}_{i_1}\ldots \widetilde{\op{s}}_{i_n}\right| s \right\rangle
 \;,
 \end{eqnarray}
 where
 \begin{eqnarray}
\label{A3a}
    \widetilde{\op{s}}_{i_\ell} &:=& U^\ast(\Omega)\, \widehat{\op{s}}_{i_\ell}\,U(\Omega)\\
    \label{A3b}
    &=& \sum_{\nu\in\{+,-,3\}} {s}_{{i_\ell},\nu}(\Omega)\,\widehat{\op{s}}_\nu
    \;,
 \end{eqnarray}
 for $\ell=1,\ldots,n$ and the ${s}_{{i_\ell},\nu}(\Omega)$ having been defined in (\ref{D3}).
 Upon repeatedly inserting ${\mathbbm 1} = \sum_{m=-s}^s \left| m \right\rangle\left\langle m \right|$ we may rewrite
 $g({\sf M})$ as
 \begin{equation}\label{A4}
  g({\sf M})= \sum_{m_1,\ldots,m_{n-1}}\left\langle s \left|  \widetilde{\op{s}}_{i_1}\right| m_1\right\rangle
  \left\langle m_1 \left|  \widetilde{\op{s}}_{i_2}\right| m_2\right\rangle
  \ldots
  \left\langle m_{n-1} \left|  \widetilde{\op{s}}_{i_n}\right| s\right\rangle
  \;.
 \end{equation}
Since all $\widetilde{\op{s}}_{i_\ell}$ are tridiagonal matrices the set of all $(m_0:=s,m_1,\ldots,m_{n-1},m_n:=s)$  leading
to non-vanishing matrix elements can be restricted to those cases where the absolute value of the
difference between adjacent magnetic quantum numbers is at most one. Writing $m_\ell=: s-a_\ell$ for all $\ell=0,\ldots,n$
we may restrict the sum in (\ref{A4}) to the set of ``spin walks" ${\mathcal W}_n$. A {\em spin walk} $w\in{\mathcal W}_n$
of length $n$ is defined as a sequence of integers
$w=(a_0,\ldots,a_n)$ subject to the conditions
\begin{eqnarray}
\label{A5a}
 && 0\le a_\ell\le 2s \;,\\
 \label{A5b}
 &&  a_0 =a_n=0\;, \\
 \label{A5c}
 && \left| a_{\ell+1}-a_\ell \right| \le 1
 \;.
\end{eqnarray}
for all $\ell=0,\ldots n-1$. For example, ${\mathcal W}_2=\{ (0,0,0),(0,1,0)\}$. The size of ${\mathcal W}_n$ grows rapidly
with $n$, see Table \ref{tab1}.
\begin{table}
  \centering
 \begin{tabular}{|r|r|}
   \hline
  Length $n$ & Number of walks\\
  \hline\hline
   1&1 \\
    \hline
   2&2 \\
      \hline
   3&4\\
      \hline
   4&9 \\
      \hline
   5&21 \\
      \hline
   6&51 \\
      \hline
   7&127 \\
      \hline
   8&323 \\
      \hline
   9&835\\
      \hline
   10&2188\\
   \hline
 \end{tabular}
  \caption{Number of spin walks of length $n$, without considering the restriction
  $a_\ell\le 2s$ in (\ref{A5a}).}\label{tab1}
\end{table}

For each walk $w=(a_0,\ldots,a_n)\in{\mathcal W}_n$ we define the sequence of ``steps" by
\begin{equation}\label{A6}
 \delta^{(w)}(\ell):= a_{\ell} - a_{\ell-1}\quad \mbox{for }\ell=1,\ldots, n+1
 \;.
\end{equation}

Note that there are three possibilities for the matrix elements
$\left\langle m_{\ell-1} \left|  \widetilde{\op{s}}_{i_{\ell}}\right| m_{\ell}\right\rangle$ in (\ref{A4}):
\begin{itemize}
  \item Either $m_{\ell-1}=m_{\ell}$ and hence $\delta^{(w)}(\ell)=0$, and the matrix $\widetilde{\op{s}}_{i_{\ell}}$ contributes
  only by the term $s_{i_{\ell},3}(\Omega)\,\widehat{\op{s}}_3$,
  \item or $m_{\ell-1}=m_{\ell}-1$ and hence $\delta^{(w)}(\ell)=-1$, and the matrix $\widetilde{\op{s}}_{i_{\ell}}$ contributes
  only by the term $s_{i_{\ell},-}(\Omega)\,\widehat{\op{s}}_-$,
  \item or $m_{\ell-1}=m_{\ell}+1$ and hence $\delta^{(w)}(\ell)=+1$, and the matrix $\widetilde{\op{s}}_{i_{\ell}}$ contributes
  only by the term $s_{i_{\ell},+}(\Omega)\,\widehat{\op{s}}_+$.
\end{itemize}
In each of these cases the matrix element can be factored into the function $s_{i_{\ell},\nu}(\Omega)$ and the remaining
matrix element of $\widehat{\op{s}}_\nu$. Then $g({\sf M})$ assumes the form
\begin{equation}\label{A7}
 g({\sf M}) = \sum_{w\in{\mathcal W}_n} N(w) \prod_{\ell=1}^{n}{s}_{{i_\ell},\delta^{(w)}(\ell)}(\Omega)
 \;,
\end{equation}
where we have conveniently identified the index set $\{+,-,3\}$ for the second index of ${s}_{{i_\ell},\nu}$ with the set
$\{ +1,-1,0\}$. The factor $N(w)$ represents the product of the matrix elements of the $\widehat{\op{s}}_{\delta^{(w)}(\ell)}$
and is independent of the $ \widetilde{\op{s}}_{i_{\ell}}$. It can be calculated as follows.

Due to the scaling of the $\widehat{\op{s}}_\nu$ a common factor $s^{-n}$ of $N(w)$ can be pulled out.
Further, the walk $w=(a_0,\ldots,a_n)$ can be divided into ``transitions" where $a_{\ell-1}\neq a_\ell$ and ``breaks"
where $a_{\ell-1}= a_\ell=:\tau$. Each break contributes the matrix element
$\left\langle s-a_{\ell-1}\left| \op{s}_3 \right| s-a_\ell\right\rangle=s-\tau$. If there are $B(\tau)$ breaks at the value $\tau$
the corresponding contribution to $N(w)$ amounts to
\begin{equation}\label{A8}
  \prod_{\tau=0}^{2s}(s-\tau)^{B(\tau)}
  \;.
\end{equation}
Next consider transitions. Each transition must be passed in both directions since the walk starts at $a_0=0$ and ends at $a_n=0$.
Hence the set of transitions can be further
divided into pairs of the form $\sigma:=a_{\ell-1} \rightarrow \sigma+1 = a_{\ell}$ and $\sigma+1= a_{k-1} \rightarrow \sigma=a_k$.
Let $T(\sigma)$ be the number of such pairs of transitions $\sigma \leftrightarrow \sigma+1$.
The corresponding matrix elements will be
\begin{equation}\label{A9}
  \left\langle s-a_{\ell-1}\left|\op{s}_+ \right| s-a_\ell\right\rangle
 \stackrel{(\ref{Dspin})}{=}
  \sqrt{s(s+1) -(s-\sigma-1)(s-\sigma)}=\sqrt{(2s-\sigma)(\sigma+1)}
  \;,
\end{equation}
and
\begin{equation}\label{A10}
  \left\langle s-a_{k-1}\left|\op{s}_- \right| s-a_k\right\rangle
  \stackrel{(\ref{Dspin})}{=}
  \sqrt{s(s+1) -(s-\sigma)(s-\sigma-1)}=\sqrt{(2s-\sigma)(\sigma+1)}
  \;,
\end{equation}
the same as (\ref{A9}). Hence the total contribution of the breaks to $N(w)$ amounts to
\begin{equation}\label{A11}
 \prod_{\sigma=0}^{2s}\left( (2s-\sigma)(\sigma+1)\right)^{T(\sigma)}
 \;,
\end{equation}
and, summarizing,
\begin{equation}\label{A12}
 N(w)=\frac{1}{s^n}\, \prod_{\tau=0}^{2s}(s-\tau)^{B(\tau)}\;\prod_{\sigma=0}^{2s}\left(  (2s-\sigma)(\sigma+1)\right)^{T(\sigma)}
 \;.
\end{equation}

It is obvious that the $s$-dependence of $g({\sf M})$ is only contained in the factor $N(w)$.
The walk $w=(0,0,\ldots,0)$ consists only of a break of length $B(0)=n$ and hence $N(w)=1$. This is the only walk
that yields a contribution to $g({\sf M})$ independent of $s$ and hence represents the classical limit
\begin{equation}\label{A12a}
{\sf M}^{\text{(cl)}}:=\Omega_{i_1}\ldots \Omega_{i_n}
\end{equation}
of $g({\sf M})$.
Each pair of transitions gives a contribution of order $O(s)$
which has to be multiplied by the corresponding prefactor $s^{-2}$ resulting in an $O(s^{-1})$-term.
Hence the semi-classical corrections to $g({\sf M})$  of lowest order $O(s^{-1})$
are obtained by those walks that contain {\em exactly} one pair of transitions.
This can be used to calculate these corrections, written as the sum of three contributions $C_1+C_2+C_3$
as follows.

\subsection{Semiclassical corrections}
\label{sec:SC}

We consider a general walk with exactly two transitions at $a_{k-1}=0 \rightarrow a_k=1$ and $a_{\ell-1}=1 \rightarrow a_\ell=0$.
The product of the corresponding two matrix elements is
\begin{equation}\label{A13}
 p(k,\ell):= \left\langle s \left| \widetilde{\op{s}}_k \right| s-1\right\rangle\,
  \left\langle s-1 \left| \widetilde{\op{s}}_\ell \right| s\right\rangle
  \;.
\end{equation}
This product contributes to $g({\sf M})$ with an $O(s^{-1})$-factor whereas the remaining matrix elements only contribute
$O(1)$-factors that can be expressed by the classical limit of the corresponding factors of the monomial ${\sf M}$.
First, we concentrate on the real part of the semi-classical corrections.
By means of (\ref{D4a}) - (\ref{D4c}) we can evaluate the product (\ref{A13}) and obtain
\begin{equation}\label{A14}
 \text{Re} \left( p(k,\ell)\right)= \frac{2}{s}\,\frac{1}{4}\, R_{k,\ell}(\Omega)
 \;,
\end{equation}
where $R(\Omega)$ is the $3\times 3$-matrix
\begin{equation}\label{A15}
 R(\Omega):= \left(
 \begin{array}{ccc}
   1-x^2 &-x\,y &-x\,z \\
   -x\,y & 1-y^2 & -y\,z \\
   -x\,z & -y\,z & 1-z^2
 \end{array}
\right)
\;.
\end{equation}
Here have rewritten the components (\ref{D1}) of $\Omega$ in the more concise form
\begin{equation}\label{AB14}
 \Omega=\left(
 \begin{array}{l}
   \sin \theta\, \cos\varphi \\
   \sin \theta\, \sin\varphi \\
   \cos\theta
 \end{array}
 \right)
 =\left(
 \begin{array}{l}
  x \\
  y \\
  z
 \end{array}
 \right)
 \;,
\end{equation}
although this representation is not unique due to $x^2+y^2+z^2=1$. For example, the classical limit
${\sf M}^{\text{(cl)}}$
of $g\left(\widehat{\op{s}}_1\, \widehat{\op{s}}_2\, \widehat{\op{s}}_1\, \widehat{\op{s}}_3\right)$
will be $x^2\,y\,z$ corresponding to the walk $(0,0,0,0,0)$.

First, we ignore the identity matrix contained in $R(\Omega)$ and note that remaining matrix elements of
 $R(\Omega)$ equal the classical factors $\Omega_k \Omega_\ell$ up to a sign. Together with the remaining classical factors
 we obtain a first contribution $C_1$ to the semi-classical correction of the form
 \begin{equation}\label{A16}
  C_1=(-1)\, \frac{2}{s}\,\frac{1}{4}{n \choose 2} {\sf M}^{\text{(cl)}}= - \frac{n(n-1)}{4\,s}\,{\sf M}^{\text{(cl)}}
 \;,
 \end{equation}
 where we have used that there are  ${n \choose 2}$ walks with exactly one transition pair.

The identity matrix part of  $R(\Omega)$ will be effective if there are repeated factors in the monomial ${\sf M}$.
In that case we obtain again ${\sf M}^{\text{(cl)}}$ but with one pair of repeated factors deleted.
It can be easily shown that the corresponding contribution  $C_2$ to the semi-classical correction can
be written as
\begin{equation}\label{A17}
C_2=\frac{1}{4\,s} \Delta\,{\sf M}^{\text{(cl)}}
\;,
\end{equation}
where $\Delta$ denotes the Laplacian $\Delta:= \frac{\partial^2}{\partial x^2}+ \frac{\partial^2}{\partial y^2}+ \frac{\partial^2}{\partial z^2}$
and ${\sf M}^{\text{(cl)}}$ is written as a monomial in the variables $x,y,z$.

For the imaginary part of the semi-classical correction of order $O(s^{-1})$ it will suffice to note that
an analogous procedure as above can be performed but with the symmetric matrix (\ref{A15}) being replaced by the anti-symmetric one
\begin{equation}\label{A18}
  I(\Omega):= \left(
 \begin{array}{ccc}
   0 &z &-y \\
   -z & 0 & x \\
   y & -x & 0
 \end{array}
\right)
\;.
\end{equation}
This implies that the last contribution $C_3$ can be written as
\begin{equation}\label{A19}
C_3 = \frac{1}{2} \,{\sf C}\left({\sf M} \right)
\;,
\end{equation}
where the operation ${\sf C}$ on ${\sf M}$ yields a classical monomial of degree reduced by one.
It is defined by summing over all ${n\choose 2}$ pairs of arguments $\widehat{\op{s}}_k,\widehat{\op{s}}_\ell$ of ${\sf M}$,
replacing each pair by its commutator $\left[\widehat{\op{s}}_k,\widehat{\op{s}}_\ell\right]$, thereby producing
a factor $\frac{\sf i}{s}$, and finally replacing all scaled spin operators by their classical limits $x,y,z$.
We will give an example in the next subsection.

\subsection{Example}
\label{sec:AB}

We will consider an example with $n=4$ and
\begin{equation}\label{AB13}
 {\sf M} = \widehat{\op{s}}_1\, \widehat{\op{s}}_2\, \widehat{\op{s}}_1\, \widehat{\op{s}}_3
 \;.
\end{equation}
There are $9$ walks $w\in {\mathcal W}_4$ with corresponding contributions to $g({\sf M})$ according to
Table \ref{tab2}.

\begin{table}
  \centering
 \begin{tabular}{|r|r|}
   \hline
  Walk $w$ & Contribution to $g({\sf M})$\\
  \hline\hline
   $(0,0,0,0,0)$ & $x^2 y z$ \\
    \hline
   $(0,1,0,0,0)$&$\frac{2}{s}\left( \frac{1}{4}({\sf i}\,z -x\,y)\,x z\right)$ \\
      \hline
   $(0,0,1,0,0)$&$\frac{2}{s}\left( \frac{1}{4}(-{\sf i}\,z -x\,y)\,x z\right)$\\
      \hline
   $(0,0,0,1,0)$&$\frac{2}{s}\left( \frac{1}{4}(-{\sf i}\,y -x\,z)\,x y\right)$ \\
      \hline
   $(0,1,1,0,0)$&$\frac{2(s-1)}{s^2}\left( \frac{1}{4}(1-x^2)\,y \,z\right)$ \\
      \hline
  $(0,1,0,1,0)$&$\frac{4}{s^2}\left( \frac{1}{16}\left({\sf i}\,x (y^2-z^2) +(1+x^2)\,y\,z\right)\right)$ \\
      \hline
  $(0,0,1,1,0)$&$\frac{2(s-1)}{s^2}\left( \frac{1}{4}({\sf i}\,x-y\,z)\,x^2)\right)$ \\
      \hline
  $(0,1,1,1,0)$&$\frac{2(s-1)^2}{s^3}\left( -\frac{1}{4}({\sf i}\,y+\,x\,z)\,x\,y)\right)$ \\
      \hline
  $(0,1,2,1,0)$&$\frac{4(2s-1)}{s^3}\left(\frac{1}{16}(1-x^2) ({\sf i}\, x -y\,z) \right)$\\
      \hline
 \end{tabular}
  \caption{Contributions to  $g({\sf M})$ for the example (\ref{AB13})
  according to the $9$ walks $w\in{\mathcal W}_4$.}\label{tab2}
\end{table}

We will explain the procedure to obtain the results of Table \ref{tab2} for the last walk
$w=(0,1,2,1,0)$ that is, however, only possible for $s\ge 1$ due to (\ref{A5a}).
We have two transition pairs $0 \leftrightarrow 1$ and $1 \leftrightarrow 2$.
The contribution
of $w$ to $g({\sf M})$ is hence obtained as
\begin{eqnarray}
\label{AB15a}
  && \left\langle s \left|\widetilde{\op{s}}_1\right| s-1 \right\rangle\,
  \left\langle s-1 \left|\widetilde{\op{s}}_2\right| s-2 \right\rangle\,
  \left\langle s-2 \left|\widetilde{\op{s}}_1\right| s-1 \right\rangle\,
  \left\langle s-1 \left|\widetilde{\op{s}}_3\right| s \right\rangle\\
  \label{AB15b}
&=& \frac{2}{s}\,\frac{2(2s-1)}{s^2}\,\widetilde{s}_{1,+}(\Omega)\,
\widetilde{s}_{2,+}(\Omega)\,\widetilde{s}_{1,-}(\Omega)\,\widetilde{s}_{3,-}(\Omega)\\
\label{AB15c}
&=& \frac{4(2s-1)}{s^3}\left(\frac{1}{16}(1-x^2) ({\sf i}\, x -y\,z) \right)
\;,
\end{eqnarray}
where the result in (\ref{AB15c}) has been obtained via (\ref{D4a}) and (\ref{D4b}) and some transformations
in the variables $x,y,z$. We note that this contribution is of order $O(s^{-2})$ due to the presence of {\em two} transition pairs.

Summing all contributions according to Table \ref{tab2} gives the final result
\begin{eqnarray}
\nonumber
 g({\sf M}) &=& x^2 y z+\frac{{(1-6 x^2) y z+\sf i}x( x^2-2 y^2)}{2 s} +\\
 \label{AB16}
   &&+\frac{(11  x^2-3) y z-{\sf i} x \left(4 x^2-5 y^2+z^2-2\right)}{4 s^2} +
    \frac{\left(1-3 x^2\right) y z+{\sf i} x \left(x^2-2 y^2-1\right)}{4 s^3}
 \;,
\end{eqnarray}
that has been ordered by inverse powers of $s$. It is a complex function since ${\sf M}$ is not Hermitean.

We will check the parts $C_1,C_2,C_3$, considered in the preceding subsection, of the lowest order semi-classical correction
\begin{equation}\label{AB17}
C:= \frac{{(1-6 x^2) y z+\sf i}x( x^2-2 y^2)}{2 s}
\end{equation}
of (\ref{AB16}). First,
\begin{equation}\label{AB18}
C_1\stackrel{(\ref{A16})}{=}  - \frac{n(n-1)}{4\,s}\,{\sf M}^{\text{(cl)}} = -\frac{3}{s}\,x^2 y z
\;,
\end{equation}
which correctly equals the monomial part of $C$ of degree $4$. Second, the remaining
part of  $\text{Re}(C)$ can be written as
\begin{equation}\label{AB19}
C_2\stackrel{(\ref{A17})}{=} \frac{1}{4\,s} \Delta\,{\sf M}^{\text{(cl)}} = \frac{1}{4\,s} \Delta\,x^2\, y\, z
= \frac{1}{2\,s}\,y\,z
\;.
\end{equation}
Finally, the imaginary part $C_3$ of $C$  can be obtained by the following self-explaining calculation
\begin{eqnarray}\label{AB20}
x\,y\,x\,z &\mapsto& {\sf C} \left(x\,y\,x\,z\right) =
\left[ x\,y\right]\, x\,z +
\left[ x\,y\,x\right]\,z + \left[ x\,y\,x\,z\right]+ x\,\left[ y\,x\right]\,z +x\,\left[ y\,x\,z\right]+x\,y\,\left[ x\,z\right]\\
&=& z\,x\,z+ 0 - y\,x\,y -x\,z\,z +x\,x\,x - x\,y\,y = x^3 -2\, x\,y^2= x\,(x^2-2\,y^2)
\;,
\end{eqnarray}
where multiplication with  $\frac{\sf i}{2\,s}$ is understood and
we have symbolically written the commutation relations of the spin operators as $\left[ x\, u \,y\right] =u\, z$ etc.~.

\section{Spin coherent states revisited}
\label{sec:CS}

\subsection{Case of single spin}
\label{sec:SS}

We add some considerations on the theory of coherent spin states,
which are not strictly necessary for the proof of the main result in Section \ref{sec:G},
but which may be useful for the understanding the classical limit of spin systems.

First, we consider a single spin with Hilbert space ${\mathcal H}^{(1,s)}\cong {\mathbbm C}^{2s+1}$
and the $(2s+1)$-dimensional, unitary irrep $g\in SU(2) \mapsto U_g$  
generated by the spin operators $\mathbf{\op{s_i}},\;i=1,2,3$ as in Section \ref{sec:D}.
We may view the set $SU(2)^{(s)}:=\{U_g \left|\right.g\in SU(2)\}$ as a matrix Lie group, isomorphic to $SU(2)$, and the linear span of the 
$(2s+1)\times (2s+1)$-matrices ${\sf i}\,\mathbf{\op{s_i}},\;i=1,2,3,$ as its Lie algebra, denoted by $su(2)^{(s)}$.
The introduction of the imaginary unit ${\sf i}$ enables one to identify the Lie bracket of $su(2)^{(s)}$ with the commutator
of anti-hermitean matrices.

The irrep $U_g$ induces another one, denoted by $R_g$, which is essentially the adjoint representation of  $SU(2)^{(s)}$,
i.~e.~ $X\mapsto U_g\,X\,U_g^\ast$ for $X\in su(2)^{(s)}$. Upon fixing the basis $\left({\sf i}\,\mathbf{\op{s_i}}\right)_{i=1,2,3}$
in  $su(2)^{(s)}$ we may write the arbitrary element $X\in su(2)^{(s)}$ as 
$X={\sf i}\,{\mathbf x}\cdot \mathbf{\op{s}}={\sf i}\,\sum_{i=1}^3 x_i\,\mathbf{\op{s_i}}$. It follows that 
$Y:=U_g\,X\,U_g^\ast= {\sf i}\,\mathbf{y}\cdot \mathbf{\op{s}}$ and $R_g$ may be defined by means of the following equation
\begin{equation}\label{CS1}
U_g \,\left({\mathbf x}\cdot \mathbf{\op{s}}\right)\,U_g^\ast = {\mathbf y}\cdot \mathbf{\op{s}}= \left( R_g\,{\mathbf x} \right)\cdot\mathbf{\op{s}}
\;,
\end{equation}
(omitting the ${\sf i}$'s) that holds for all ${\mathbf x}\in{\mathbbm R}^3$. 

Upon the identification $su(2)^{(s)}\cong {\mathbbm R}^3$, the map $g\mapsto R_g$  is an adjoint representation, 
and hence $R_g$ will linear and invertible. Moreover, the commutation relations for
the $\mathbf{\op{s_i}},\;i=1,2,3$ imply 
\begin{equation}\label{commrel}
\left[{\mathbf a}\cdot \mathbf{\op{s}},{\mathbf b}\cdot \mathbf{\op{s}}\right]={\sf i}\,({\mathbf a}\times{\mathbf b})\cdot\mathbf{\op{s}}
\end{equation}
for all ${\mathbf a},{\mathbf b}\in{\mathbbm R}^3$. Hence $R_g$ is a Lie algebra homomorphism of $\left({\mathbbm R}^3, \times \right)$.
From this it easily follows that $R_g$  leaves the scalar product and the scalar triple product of vectors invariant and hence $R_g\in SO(3)$:

\begin{lemma}\label{L4}
 If $A: {\mathbbm R}^3 \rightarrow {\mathbbm R}^3$ is a linear and invertible map satisfying
 $A({\mathbf a}\times {\mathbf b})=\left(A\,{\mathbf a} \right)\times \left(A\,{\mathbf b} \right)$ 
 for all ${\mathbf a},{\mathbf b}\in {\mathbbm R}^3$, then also 
 $A({\mathbf a})\cdot A({\mathbf b})={\mathbf a}\cdot {\mathbf b}$
 holds.
\end{lemma}
{\bf Proof}: We may assume ${\mathbf b}\neq {\mathbf 0}$ and choose ${\mathbf c} \in {\mathbbm R}^3$
such that $\{ {\mathbf b},{\mathbf c} \}$ is linearly independent. Since $A$ is invertible, also 
$\{A\, {\mathbf b},A\,{\mathbf c} \}$  will be linearly independent.
Applying $A$ to the l.~h.~s.~of the triple product expansion
\begin{equation}\label{bac}
 {\mathbf a}\times \left({\mathbf b}\times {\mathbf c} \right)= ({\mathbf a}\cdot  {\mathbf c})\, {\mathbf b} - ({\mathbf a}\cdot  {\mathbf b})\, {\mathbf c}
\end{equation}
yields
\begin{equation}\label{baclhs}
A\,\left( {\mathbf a}\times \left({\mathbf b}\times {\mathbf c} \right) \right)=
(A\,{\mathbf a})\times \left((A,{\mathbf b})\times (A\,{\mathbf c}) \right)= 
((A\,{\mathbf a})\cdot (A\,{\mathbf c}))\, (A\,{\mathbf b}) - ((A\,{\mathbf a})\cdot  (A\,{\mathbf b}))\, (A\,{\mathbf c})
\;.
\end{equation}
On the other hand, applying $A$ to the r.~h.~s.~of (\ref{bac}) and using that $A$ is linear, yields
\begin{equation}\label{bacrhs}
A\,\left(  ({\mathbf a}\cdot  {\mathbf c})\, {\mathbf b} - ({\mathbf a}\cdot  {\mathbf b})\, {\mathbf c}\right)=
({\mathbf a}\cdot  {\mathbf c})\, (A\,{\mathbf b}) - ({\mathbf a}\cdot  {\mathbf b})\,(A\, {\mathbf c})
\;.
\end{equation}
Since $\{A\, {\mathbf b},A\,{\mathbf c} \}$ is linearly independent, both expressions, (\ref{baclhs}) and  (\ref{bacrhs}), 
can only be equal if  $A({\mathbf a})\cdot A({\mathbf c})={\mathbf a}\cdot {\mathbf c}$ and  $A({\mathbf a})\cdot A({\mathbf b})={\mathbf a}\cdot {\mathbf b}$.
\hfill$\Box$\\

We note without proof that $R:SU(2)\rightarrow SO(3),\, g\mapsto R_g$, is an irreducible representation independent of the spin quantum number $s$
and can be viewed as the universal covering homomorphism of $SO(3)$.

The special choice ${\mathbf x}={\scriptsize{\mathbf e}_3:=\left( \begin{array}{c}
                                                                0 \\
                                                                0 \\
                                                                1
                                                              \end{array} \right)}$ yields
\begin{equation}\label{CS2}
U_g\,{\mathbf e}_3 \cdot \op{s}\;U_g^\ast
=U_g\,\op{\mathbf s}_3 \,U_g^\ast = \Omega\cdot \op{s}
\;,
\end{equation}
where
\begin{equation}\label{CS3}
 \Omega =R_g\, {\mathbf e}_3\in{\mathbbm S}^2
\end{equation}
can be arbitrarily chosen within ${\mathbbm S}^2$ . For fixed $\Omega\in{\mathbbm S}^2$ the equation (\ref{CS3}) does not uniquely
determine $g\in SU(2)$. Two possibly  different $g,h\in SU(2)$ determine the same $\Omega$ iff $g^{-1}\,h \in N_3$,
where $N_3\subset SU(2)$ denotes the subgroup $N_3\cong U(1)$ of special, unitary matrices $u$ commuting with the Pauli matrix $\op{\sigma}_3$ ,
i.~e., being of the form
\begin{equation}\label{CS4}
 u=\left( \begin{array}{cc}
            e^{{\sf i} \alpha} & 0 \\
            0 & e^{-{\sf i} \alpha}
          \end{array}
 \right)
 \;.
\end{equation}
This means that the unit vectors $\Omega\in {\mathbbm S}^2$ correspond in a $1:1$ manner to the left cosets in $SU(2)/N_3$,
which is essentially the Hopf fibration \cite{H31}.

Spin coherent states $\left| \Omega \right\rangle$ are defined as (normalized) eigenvectors of $\Omega\cdot\op{s}$ corresponding
to the largest eigenvalue $s$. This determines $\left| \Omega \right\rangle$ up to a phase. The phase convention chosen in
(\ref{D5}) that follows \cite{L73} has the advantage that the resulting dependence on the azimuthal angle $\phi$ will be of the
form $\exp\left( {\sf i} (s-m)\phi \right)$, where $s-m$ is an integer and hence will be smooth for
all points of ${\mathbbm S}^2$ except for the south pole where $\theta=\pi$ and $\phi$ cannot be defined smoothly.
For $\theta\rightarrow \pi$ we obtain $\left(\cos\theta/2\right)^{s+m}\rightarrow 1$ for $m=-s$ in (\ref{D5}) but due to
\begin{equation}\label{CS5}
 \left| \Omega \right\rangle \rightarrow \exp(2\,{\sf i}\,s \,\phi)\,\left| -s\right\rangle
 \;,
\end{equation}
the limit of $\left|\Omega\right\rangle$ for $\theta\rightarrow \pi$  is ill-defined.

This problem cannot be avoided and is due to the fact that a smooth cross section $\sigma: {\mathcal S}^2 \rightarrow SU(2)$ of the Hopf fibration
$\pi: SU(2) \rightarrow SU(2)/U(1) \cong {\mathcal S}^2$ cannot be defined globally, see also \cite{P72}.
However, the one-dimensional projections $\left| \Omega\right\rangle\left\langle \Omega \right|$
depend smoothly on $\Omega$ for all $\Omega\in{\mathbbm S}^2$ since the oscillating $\phi$-components are more and more damped
if $\Omega$ approaches the south pole.

Recall that $U_g\, \square\, U_g^\ast$ maps $\Omega\cdot \op{s}$ onto $R_g\,\Omega\cdot \op{s}$, see (\ref{CS1}).
Hence $U_g$ maps the eigenvector of $\Omega\cdot \op{s}$ corresponding to the largest eigenvalue $s$ onto
the eigenvector of $\left(R_g\,\Omega\right)\cdot \op{s}$ corresponding to the largest eigenvalue $s$, and thus
\begin{equation}\label{CS6}
  U_g\,\left| \Omega\right\rangle = e^{{\sf i} \alpha(g)}\, \left|R_g\, \Omega\right\rangle
  \;,
\end{equation}
where we have inserted the phase factor $e^{{\sf i} \alpha(g)}$ since the eigenvectors are only determined up to a phase.
Consequently, this phase factor cancels for the projections and we have
\begin{equation}\label{CS7}
 U_g\,\left| \Omega\right\rangle \left\langle \Omega \right| \,U_g^\ast =
 \left| R_g\, \Omega\right\rangle \left\langle  R_g\,\Omega \right|
 \;.
\end{equation}
As an application we note that
\begin{equation}\label{CS8}
 U_g\left(\int \left| \Omega\right\rangle \left\langle \Omega \right|\,d\Omega\right) U_g^\ast=
 \int \left|R_g\, \Omega\right\rangle \left\langle R_g\Omega \right|\,d\Omega=
 \int \left| \Omega\right\rangle \left\langle \Omega \right|\,d\Omega
 \;,
\end{equation}
using that the surface element $d\Omega$ is invariant under rotations. It follows that
$ \int \left| \Omega\right\rangle \left\langle \Omega \right|\,d\Omega$ commutes with all $U_g$ and hence
must be a multiple of the unit operator, $U_g$ being irreducible. This yields an alternative proof of
the completeness relation (\ref{Did}), if the remaining multiple is determined by taking the trace
of both sides of (\ref{Did}).

\subsection{Case of $N$ spins}
\label{sec:SN}
Next we consider $N$ spins and largely retain the notations of Section \ref{sec:D}.
The Hilbert space
of the $N$ spin system ${\mathcal H}^{(N,s)}$ will be simply denoted by ${\mathcal H}$. Moreover,
$L^2$ denotes the Hilbert space
\begin{equation}\label{CS9}
  L^2:=L^2\left( {\mathcal S}, \widetilde{d}\Omega\right)
  \;, \mbox{where }\widetilde{d}\Omega:=\left(\frac{2s+1}{4\pi}\right) d\Omega_1\ldots d\Omega_N
  \;.
\end{equation}
Sometimes different scalar products will be distinguished by subscripts referring to the two different Hilbert spaces.
We suppress the $s$-dependence of various quantities since $s$ will be arbitrary but kept fixed in this subsection.
We do not longer use boldface symbols in this subsection. Hence the completeness relation (\ref{DidN}) reads
\begin{equation}\label{CS10}
  {\mathbbm 1}_{\mathcal H}=\int \left| \Omega\right\rangle \left\langle \Omega \right|\,\widetilde{d}\Omega
  \;.
\end{equation}
We note in passing that the ``resolution of the identity" (\ref{CS10}) gives rise to a generalized observable \cite{BLM96}
in the sense of a covariant POV measure $F$ defined on Borel subsets ${\mathcal A}$  of ${\mathcal S}$  by means of
\begin{equation}\label{CS11}
  F\left( {\mathcal A}\right):= \int_{\mathcal A} \left| \Omega\right\rangle \left\langle \Omega \right|\,\widetilde{d}\Omega
  \;.
\end{equation}
$F$ physically describes unsharp joint measurements of the (non-commuting) spin operators $\op{s}_i,\;i=1,2,3$.

Next we define a linear map
\begin{eqnarray}\nonumber
 && j:{\mathcal H}\rightarrow L^2\\
\label{CS12}
&& j(\varphi)(\Omega):= \left\langle \Omega | \varphi \right\rangle
\;,
\end{eqnarray}
that turns out to be a partial isometry, i.~e., satisfying
\begin{equation}\label{CS13}
  j^\ast\,j={\mathbbm 1}_{\mathcal H}
  \;.
\end{equation}
{\bf Proof} of (\ref{CS13}):
For any $\varphi, \psi \in{\mathcal H}$ we have
\begin{eqnarray}
\label{CS14a}
  \left\langle \varphi \left|j^\ast\,j \right|\psi \right\rangle_{\mathcal H}
  &=&
  \left\langle j\, \varphi \left|j \right.\psi \right\rangle_{L^2} \\
  \label{CS14b}
   &=& \int \overline{(j\,\varphi)(\Omega)}\,(j\,\psi)(\Omega)\,\widetilde{d}\Omega\\
   \label{CS14c}
   &\stackrel{(\ref{CS12})}{=}&\int \langle\varphi| \Omega\rangle_{\mathcal H}\, \langle \Omega | \psi\rangle_{\mathcal H}\,\widetilde{d}\Omega\\
   \label{CS14d}
   &=& \left\langle \varphi \left| \int \left| \Omega\right\rangle \left\langle \Omega \right|\,\widetilde{d}\Omega \right| \psi \right\rangle_{\mathcal H}\\
   \label{CS14e}
   &\stackrel{(\ref{CS10})}{=}& \left\langle \varphi \left| {\mathbbm 1}_{\mathcal H}   \right| \psi \right\rangle_{\mathcal H}
   \;.
\end{eqnarray}
Since $\varphi,\,\psi\in{\mathcal H}$ are arbitrary this concludes the proof. \hfill$\Box$\\

It follows that ${\mathbbm P}:= j\,j^\ast$ will be a projector in $L^2$. ${\mathbbm P}$ can be  written as an integral operator
with kernel $\left\langle \Omega' \left| \right. \Omega\right\rangle$. The relation ${\mathbbm P}^2={\mathbbm P}$
when written in terms of integral operators is often referred to as the ``reproducing kernel property", see, e.~g.,
\cite{L73}, (2.18) for $N=1$.

Next we consider a complex function $F:{\mathcal S}\rightarrow {\mathbbm C}$ such that
$\Omega\mapsto \left|F(\Omega)\right|$ is integrable and denote by $\widehat{F}$ the
corresponding multiplication operator in $L^2$:
\begin{equation}\label{CS14}
  \widehat{F}(G)(\Omega):= F(\Omega)\,G(\Omega)
\end{equation}
for all $G\in L^2$. One defines the ``contravariant matrix" of $F$ by the linear operator
\begin{eqnarray}
\nonumber
  A(F) &:& {\mathcal H}\rightarrow{\mathcal H}\\
  \label{CS15}
  A(F)&{:=}& j^\ast\, \widehat{F}\, j
  \;.
\end{eqnarray}
The contravariant matrix can be expressed in a more direct way by writing, for arbitrary $\varphi,\,\psi\in{\mathcal H}$,
\begin{eqnarray}
\label{CS16a}
 \left\langle \varphi \left| A(F)\right| \psi\right\rangle_{\mathcal H}&\stackrel{(\ref{CS15})}{=}&
  \left\langle \varphi \left|j^\ast \, \widehat{F}\,j\right| \psi\right\rangle_{\mathcal H}\\
  \label{CS16b}
  &=& \left\langle j\,\varphi \left| \widehat{F}\right| j\,\psi\right\rangle_{L^2}\\
  \label{CS16c}
  &\stackrel{(\ref{CS14})}{=}&\int \overline{(j\,\varphi)(\Omega)}\,F(\Omega)\, (j\,\psi)(\Omega) \, \widetilde{d}\Omega\\
  \label{CS16d}
  &\stackrel{(\ref{CS12})}{=}&\int \langle \varphi\left| \right. \Omega\rangle_{\mathcal H}\,F(\Omega)\,
  \langle \Omega\left| \right. \psi\rangle_{\mathcal H} \, \widetilde{d}\Omega\\
  \label{CS16e}
  &=& \left\langle \varphi \left| \int F(\Omega)\,\left| \Omega\right\rangle
  \left\langle\Omega \right| \widetilde{d}\Omega \right| \psi\right\rangle_{\mathcal H}
  \;.
\end{eqnarray}
  Hence
\begin{equation}\label{CS17}
  A(F) = \int F(\Omega)\,\left| \Omega\right\rangle   \left\langle\Omega \right| \widetilde{d}\Omega
  \;,
\end{equation}
and $F$ is a ``contravariant symbol" of $A(F)$ in the sense of \cite{B75}. Our wording follows \cite{B75}
but the notation is due to \cite{L73}. As these authors emphasize, the map $F\mapsto A(F)$ is many-to-one
and hence the ``contravariant symbol of $A$", denoted by $G(A)$, is not unique, but has to be chosen in some way.
In contrast, the covariant symbol $g(A)$, defined in (\ref{Dcov}), is always unique.

Following G.~Ludwig \cite{L12}, the common language to describe quantum and classical systems involve the notions of a ``statistical duality"
consisting of ``states", ``effects" and the ``probability" that an effect will be triggered in a certain state of the system.
In the quantum mechanics of spin systems states will be mathematically represented by statistical operators $W$ on ${\mathcal H}$,
effects by Hermitean operators $A$ on  ${\mathcal H}$ with all eigenvalues $e$ satisfying $0\le e\le 1$ such that the corresponding
probability is given by
\begin{equation}\label{CS18}
  p=\mbox{Tr}\left(W\,A\right)
  \;.
\end{equation}
In the classical realm states and effects are represented by functions defined
on ${\mathcal S}$ and the probability is given by the integral over ${\mathcal S}$ of the product of these functions.
To be more precise, we represent classical states by functions $w\in L^1\left( {\mathcal S},\widetilde{d}\Omega\right)$ satisfying
$w(\Omega)\ge 0$ for all $\Omega\in {\mathcal S}$ and $\int{w(\Omega)}\,\widetilde{d}\Omega=1$, and effects by real functions
$a\in L^\infty\left( {\mathcal S},\widetilde{d}\Omega\right)$ satisfying $0\le a(\Omega)\le 1$ for all $\Omega\in {\mathcal S}$.
The probability $p$ is then given by
\begin{equation}\label{CS19}
  p=\int w(\Omega)\,a(\Omega)\,\widetilde{d}\Omega
  \;.
\end{equation}

The benefit of using the theory of spin coherent states is the observation that the quantum mechanical statistical duality
can also be brought into the form of the classical one. More specifically, we can prove the following
\begin{prop}\label{Pstat}
 Let $W:{\mathcal H}\rightarrow{\mathcal H}$ be a statistical operator and $F:{\mathcal S}\rightarrow {\mathbbm R}$ be
 a function $F\in L^\infty\left( {\mathcal S},\widetilde{d}\Omega\right)$,giving rise to a
 Hermitean operator $A(F): {\mathcal H}\rightarrow {\mathcal H}$.
 Let $g(W)$ denote the covariant symbol of $W$. Then the following holds
 \begin{equation}\label{CS20}
  \mbox{Tr}\left( W\, A(F)\right) = \int g(W)\,F(\Omega)\, \widetilde{d}\Omega
  \;.
 \end{equation}
\end{prop}
{\bf Proof}: From
\begin{equation}\label{CS21}
  A(F)\stackrel{(\ref{CS17})}{=}\int F(\Omega)\, \left| \Omega\right\rangle\left\langle\Omega\right| \,\widetilde{d}\Omega
\end{equation}
it follows that
\begin{eqnarray}
\label{CS22a}
  \mbox{Tr}\left( W\, A(F)\right) &=& \int F(\Omega)\,\mbox{Tr}\left( W\,\left| \Omega\right\rangle\left\langle\Omega\right|\right) \,\widetilde{d}\Omega\\
  &=&\int F(\Omega) \, \left\langle \Omega \left| W \right|\Omega \right\rangle\,\widetilde{d}\Omega\\
  &=& \int F(\Omega) \, g(W)\,\widetilde{d}\Omega
  \;.
\end{eqnarray}
This concludes the proof of Proposition \ref{Pstat}. \hfill $\Box$\\

In particular, it is now obvious that the classical limit of Gibbs states $G^{(N,s)}$ should be investigated in terms of
the limit $s\rightarrow\infty$ of the covariant symbols $g(G^{(N,s)})$. For the corresponding topology we have taken a pragmatic
view: Since the $g(G^{(N,s)})$ and their classical limit $G^{(\text{cl})}$ are always continuous functions ${\mathcal S}\rightarrow {\mathbbm R}$
we have proven the convergence of $g(G^{(N,s)})$ towards $G^{(\text{cl})}$ w.~r.~t.~the sup-norm which makes ${\mathcal C}({\mathcal S})$ into a Banach space.

The group $SU(2)$ operates on $B_h({\mathcal H})$ via $A\mapsto U_g\,A\,U_g^\ast$ and on $L^\infty\left( {\mathcal S},\widetilde{d}\Omega\right)$
via $a\mapsto a \circ R_g^{-1}$. We will show that $F\mapsto A(F)$ intertwines between these representations:
\begin{prop}\label{Pinter}
\begin{equation}\label{CS23}
  A\left( F\circ R_g^{-1}\right)= U_g\,A(F)\,U_g^\ast\quad
  \mbox{for all } F\in L^\infty\left( {\mathcal S},\widetilde{d}\Omega\right)
  \;.
\end{equation}
\end{prop}

\noindent
{\bf Proof}:
\begin{eqnarray}\label{CS24a}
  U_g\,A(F)\,U_g^{\ast} &\stackrel{(\ref{CS17})}{=}&
  \int F(\Omega)\, U_g\, \left|\Omega\right\rangle\left\langle\Omega\right|\,U_g^\ast \,\widetilde{d}\Omega \\
  \label{CS24b}
  &\stackrel{(\ref{CS7})}{=}&
  \int F(\Omega)\, \left| R_g\,\Omega\right\rangle\left\langle  R_g\,\Omega\right| \,\widetilde{d}\Omega \\
  \label{CS24c}
  &=& \int F(R_g^{-1}\Omega)\, \left| \Omega\right\rangle\left\langle\Omega\right| \,\widetilde{d}\Omega \\
  \label{CS24d}
  &=& A\left( F\circ R_g^{-1}\right)
  \;.
\end{eqnarray}
This concludes the proof of (\ref{CS23}). \hfill $\Box$\\

Both spaces  $B_h({\mathcal H})$ and $L^\infty\left( {\mathcal S},\widetilde{d}\Omega\right)$ can be decomposed into
irreps of $SU(2)$. For $B_h({\mathcal H})$ this is a finite number of irreps of dimensions $d=1,3,5,\ldots 4s+1$
and the corresponding subspaces $B_d$ are spanned by monomials of spin operators of degree $\frac{d-1}{2}$.
For  $L^\infty\left( {\mathcal S},\widetilde{d}\Omega\right)$ this is an infinite number of irreps of dimensions $d=1,3,5,\ldots$
and the corresponding subspaces $L_d$ are spanned by monomials of degree $\frac{d-1}{2}$ in the variables $x,y,z$.

In the following we will show that the map $F\mapsto A(F)$ is surjective,
which is implicitly assumed when defining ``contravariant symbols" but, as far as I can see, not proven.

In view of the intertwining property (\ref{CS23})
it will suffice to show that $A(L_d)\cap B_d \neq \{0\}$, i.~e., for a certain monomial ${\sf M}\in L_d$ the contravariant matrix
$A({\sf M})$ will have a non-zero component in the subspace $B_d$ of monomials of spin operators of degree  $n=\frac{d-1}{2}$.
This can be shown by calculating $A({\sf M})$ for the choice ${\sf M}=z^n$ in the representation (\ref{CS17}) and using computer-algebraic means.
The result reads as follows:
\begin{lemma}\label{Lcontra}
The contravariant matrix $A(z^n)$ of the function $a(\Omega)=z^n$ can be written as a sum
\begin{equation}\label{CS25}
  A(z^n) = \sum_{k=n,n-2,\ldots} a_k\, \op{s}_3^k
\end{equation}
with coefficients
\begin{eqnarray}
\label{CS26a}
  a_n &=& \frac{2^n\,(2 s+1)! }{(n+2 s+1)!}, \\
  \label{CS26b}
  a_{n-2}&=& \frac{n! \,2^n (2 s+1)! (1+n+3 s)}{2\times 3! (n-2)! (n+2 s+1)!},\\
  \label{CS26c}
  a_{n-4}&=& \frac{n! \,2^n (1+2 s)! (-2+n (3+5 n)+15 s (1+2 n+3 s))}{2\times 6! (n-4)! (1+n+2 s)!},\\
  \label{CS26d}
  a_{n-6}&=& \frac{n! \,2^n (2 s+1)! \left((1+n) (12+7 n (-11+5 n))+63 (-1+n) (3+5 n) s+945 n
   s^2+945 s^3\right)}{9! (n-6)! (1+n+2 s)!},\\
  \nonumber
  &\vdots&\\
  \label{CS26e}
  a_0&=& \frac{(n-1){!!} (1+2 s){!!}}{(1+n+2 s){!!}}, \quad \mbox{if } n\;\mbox{ is even}
  \;.
\end{eqnarray}
\end{lemma}

The observation that the contravariant matrix $A(z^n)$ for even $n$ will be a linear combination of even monomials $\op{s}_3^k$,
and analogously for odd $n$,
follows already from the intertwining property  (\ref{CS23}).
For our purposes of proving that  $F\mapsto A(F)$  is surjective it is only important
that the leading coefficient $a_n$ of the sum (\ref{CS25}) is non-zero. Hence we obtain:
\begin{prop}\label{Psurj}
The map $A: L^\infty\left( {\mathcal S},\widetilde{d}\Omega\right) \rightarrow B_{h}({\mathcal H})$ is surjective.
\end{prop}

\section{Summary}
\label{sec:SUM}
We have proved that the classical limit $s\to \infty$ of suitably scaled quantum Gibbs states $\op{G}(\beta)$ exists and corresponds to the classical Gibbs state with the same inverse temperature $\beta$. For this proof, we have used the representation of the quantum Gibbs states as continuous functions on the classical phase space ${\mathcal S}$ in terms of spin-coherent states and the corresponding sup-norm of ${\mathcal C}({\mathcal S})$.
The underlying Hamiltonian $\op{H}$ can be any polynomial in the spin operators.

The corresponding limit theorem for the partition function $Z(\beta)$ has already been proved in \cite{L73}.
Therefore, at first sight this result would be sufficient to obtain the classical limit of
thermodynamic functions expressible by derivatives of $Z(\beta)$.
However, although it seems plausible and has been confirmed for various examples, see, e.g., \cite{DKRS19,SS22},
it is not mathematically trivial that the limit of $Z(\beta)$ can be carried over to its derivatives,
e.~g., to the specific heat $c(\beta) =\beta^2\,\frac{\partial^2}{\partial \beta^2} \log Z$.
The present result opens an alternative way to treat the classical limit of thermodynamic functions
that can be expressed by thermal expectation values, such as
$U(\beta)=\langle \op{H} \rangle := \text{Tr}\, \left( \op{G}(\beta)\,\op{H}\right)$ or
$c(\beta) =\beta^2\left(\langle U^2 \rangle -\langle U\rangle^2 \right)$.
We recall that these expectation values can be written in terms of integrals over ${\mathcal S}$, see (\ref{CS20}).
Without going into the details, it seems clear how the corresponding theory would proceed.

\section*{Acknowledgment}

I am indebted to the members of the DFG research group FOR 2692 for continuous support and encouragement.

\end{document}